\documentclass[journal,twoside,web]{ieeecolor}
\usepackage{generic}
\usepackage{cite}
\usepackage{amsmath,amssymb,amsfonts}
\usepackage{graphicx}
\usepackage{algorithm}
\usepackage{algpseudocode}
\usepackage{hyperref}
\hypersetup{hidelinks=true}
\usepackage{textcomp}
\usepackage{booktabs}
\usepackage{pgfplots}
\usepackage[caption=false]{subfig}
\usepackage{xcolor}
\usepackage{mathtools}

\usepackage{amsthm}

\DeclareFontFamily{U}{mathx}{\hyphenchar\font45}
\DeclareFontShape{U}{mathx}{m}{n}{
      <5> <6> <7> <8> <9> <10>
      <10.95> <12> <14.4> <17.28> <20.74> <24.88>
      mathx10
      }{}
\DeclareSymbolFont{mathx}{U}{mathx}{m}{n}
\DeclareFontSubstitution{U}{mathx}{m}{n}
\DeclareMathAccent{\widecheck}{0}{mathx}{"71}
\DeclareMathAccent{\wideparen}{0}{mathx}{"75}

\DeclareMathOperator{\Reach}{\mathcal{Y}} 
\newcommand{\sys}{S}
\newcommand{\sysT}{S_{\mathrm{T}}}
\newcommand{\sysM}{S_{\mathrm{M}}}
\newcommand{\func}{\mathcal{F}}

\newcommand{\sysMu}[1]{S_{\mathrm{M}}}

\newcommand{\uncX}{\mathcal{X}_0}
\newcommand{\uncU}[1]{\mathcal{U}_{#1}}
\newcommand{\comm}[1]{\textit{\footnotesize\textcolor{TUMBlue}{#1}}}

\DeclareMathOperator*{\argmin}{arg\,min}
\allowdisplaybreaks[3] 
\algblock{Input}{EndInput}
\algnotext{EndInput}
\algblock{Output}{EndOutput}
\algnotext{EndOutput}
\newcommand{\Desc}[2]{\Statex \makebox[4em][l]{#1}#2}
\newcommand\Algphase[1]{%
\vspace{0.1cm}\Statex\textcolor{TUMBlue}{\textbf{\textit{#1}}}\vspace{0.05cm}%
}
\newcommand\copyrighttext{%
	\footnotesize This work has been submitted to the IEEE for possible publication. Copyright may be transferred without notice, after which this version may no longer be accessible.}
\newcommand\copyrightnotice{%
	\begin{tikzpicture}[remember picture,overlay]
		\node[anchor=south,yshift=10pt] at (current page.south) {\fbox{\parbox{\dimexpr\textwidth-\fboxsep-\fboxrule\relax}{\copyrighttext}}};
	\end{tikzpicture}%
}

\pgfplotsset{compat=1.8}
\usepgfplotslibrary{statistics}
\def\BibTeX{{\rm B\kern-.05em{\sc i\kern-.025em b}\kern-.08em
    T\kern-.1667em\lower.7ex\hbox{E}\kern-.125emX}}
\usepackage[capitalise]{cleveref}

\definecolor{colWhite}{RGB}{150,    204,    173}
\definecolor{colGrayId}{RGB}{34,    93,   156}
\definecolor{colBlackId}{RGB}{43,    5,    46}
\definecolor{TUMBlue}{HTML}{0065BD}
\definecolor{measBlue}{rgb}{0,    0.4470,    0.7410}
\definecolor{reachGreen}{rgb}{0.4660,    0.6740,    0.1880}
\definecolor{TUMSecondaryBlue}{HTML}{005293}
\definecolor{TUMSecondaryBlue2}{HTML}{003359}
\definecolor{TUMBlack}{HTML}{000000}
\definecolor{TUMWhite}{HTML}{FFFFFF}
\definecolor{TUMDarkGray}{HTML}{333333}
\definecolor{TUMGray}{HTML}{808080}
\definecolor{TUMLightGray}{HTML}{CCCCC6}
\definecolor{TUMAccentGray}{HTML}{DAD7CB}
\definecolor{TUMAccentOrange}{HTML}{E37222}
\definecolor{TUMAccentGreen}{HTML}{A2AD00}
\definecolor{TUMAccentLightBlue}{HTML}{98C6EA}
\definecolor{TUMAccentBlue}{HTML}{64A0C8}

\newtheorem{theorem}{Theorem}
\newtheorem{definition}{Definition}
\newtheorem{proposition}{Proposition}
\newtheorem{corollary}{Corollary}
\newtheorem{lemma}{Lemma}
\newtheorem{problem}{Problem} 
\newtheorem{subproblem}{Problem}[problem]
\crefname{equation}{}{}
\crefname{section}{Sec.}{}
\crefname{table}{Table}{}
\crefname{algorithm}{Alg.}{}
\crefname{definition}{Def.}{}
\crefname{problem}{Problem}{}
\crefname{subproblem}{Problem}{}
\crefname{theorem}{Thm.}{}
\crefname{proposition}{Prop.}{}
\crefname{lemma}{Lemma}{}

\begin{document}
\title{Reachset-Conformant System Identification}
\author{Laura Lützow and Matthias Althoff 
\thanks{
This paper was submitted for review on ...
This work was funded in part by the Deutsche Forschungsgemeinschaft (DFG, German Research Foundation) – SFB 1608 – under grant number 501798263 and in part by the European Commission project justITSELF under grant number 817629.
}
\thanks{
The authors are both with the School of Computation, Information and Technology, Technical University of Munich, 85748 Garching, Germany. \{{\tt\small laura.luetzow@tum.de}, {\tt\small althoff@in.tum.de}\} 
}}

\maketitle

\begin{abstract}
Formal verification techniques play a pivotal role in ensuring the safety of complex cyber-physical systems. To transfer model-based verification results to the real world, we require that the measurements of the target system lie in the set of reachable outputs of the corresponding model, a property we refer to as reachset conformance. This paper is on automatically identifying those reachset-conformant models. While state-of-the-art reachset-conformant identification methods focus on linear state-space models, we generalize these methods to nonlinear state-space models and linear and nonlinear input-output models. Furthermore, our identification framework adapts to different levels of prior knowledge on the system dynamics. In particular, we identify the set of model uncertainties for white-box models, the parameters and the set of model uncertainties for gray-box models, and entire reachset-conformant black-box models from data. The robustness and efficacy of our framework are demonstrated in extensive numerical experiments using simulated and real-world data.
\end{abstract}
\copyrightnotice

\begin{IEEEkeywords}
Data-defined modeling, dynamical systems, formal safety verification, input-output models, reachability analysis, reachset conformance, set-based computing, uncertainty quantification, zonotopes.
\end{IEEEkeywords}


\sloppy
\section{Introduction}
\label{sec:introduction}

Formal verification techniques require mathematical models that describe the behavior of the target system \cite{roehm2019conformance}.
Since no model of a cyber-physical system is completely accurate due to the inherent complexity of real-world dynamics, we include uncertainties in our models to cover the discrepancies to the target system.
The automatic identification of these uncertainties alongside the nominal model is the topic of this article.

\subsection{Literature Overview}

Classical system identification techniques aim to find the model or model parameters, which minimize some objective function, usually by making probabilistic assumptions about the uncertainty in the data and the model~\cite{ljung2002identification,isermann2010identification,keesman2011system,toth2010modeling}.
In many applications, however, we do not have any information about the underlying probability distributions~\cite{polyak2004ellipsoidal}. 
Furthermore, to establish rigorous safety guarantees in planning and control, it is beneficial to have models with bounded uncertainties~\cite{canale2013design}. 

\begin{figure}[t]
    \centering
    \hspace{-0.3cm}
    \includegraphics[width=\linewidth, keepaspectratio,page=2]{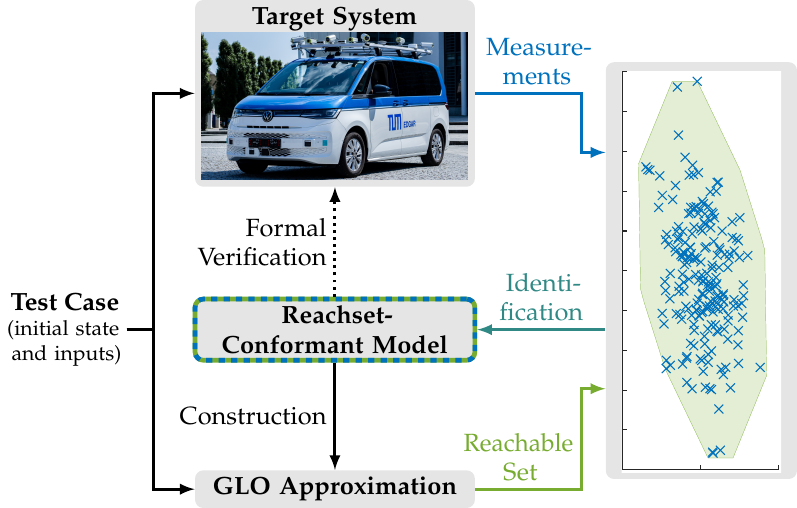}
    \caption{Identification of a reachset-conformant model: {The outputs of the target system are measured across different test cases, each defined by an initial state and an input trajectory. A reachset-conformant model is identified by adjusting its parameters and uncertainty sets such that all measurements lie within the model's reachable set, which is computed for each test case using the corresponding GLO approximation.}}
    \label{fig:id}
\end{figure}

Given the model structure and the bounds of an additive model uncertainty, which describes the errors between the model outputs and the measurements of the target system, the set of feasible parameters can be estimated by \emph{set-membership methods}, also called bounded-error approaches~\cite{walter1990survey}.
This set of feasible parameters is overapproximated by multi-dimensional intervals~\cite{kieffer2006box,polyak2005interval,casini2014feasible}, ellipsoids~\cite{polyak2004ellipsoidal}, or zonotopes~\cite{bravo2006bounded}. 
An overapproximation with arbitrary accuracy can be obtained using unions of non-overlapping multi-dimensional intervals~\cite{rauh2019interval,mahato2020validated,kieffer2011guaranteed}.
For linear systems, the exact parameter set can be represented by a polytope~\cite{mo1990polytope} if the model uncertainty is bounded by an interval, or by a matrix zonotope~\cite{alanwar2021data} in the case of additive zonotopic uncertainties. 
However, specifying the model structure and the uncertainty bounds a priori, as required for set-membership approaches, is usually a difficult task for real-world systems.
If the uncertainty bounds are assumed too large, the resulting set of possible models is also too large. 
On the other hand, we might not find any feasible parameters if the uncertainty bounds are assumed too small.

To address these problems, techniques for \emph{conformant model identification} have been developed~\cite{roehm2019conformance}. 
These methods find optimal models and their corresponding uncertainties such that a conformance relation is established. 
Depending on the conformance relation, different properties can be transferred from the conformant model to the system. 
Many publications focus on \emph{simulation relations}, which facilitates the transfer of temporal logic properties by relating system states~\cite{chen2022data,sadraddini2018formal,mulagaleti2022data}. 
Establishing a simulation relation, however, requires that the system state is observable.
Less restrictive are trace conformance and reachset conformance, which only consider the output of the system. 
In \emph{trace-conformant} identification, the model and a bounded set for the uncertainties, which we refer to as uncertainty set, are constructed such that the model is able to produce all output traces of the target system~\cite{elguindy2016formal,schuermann2017ensuring,kochdumper2022conformant}.
In most cases, this can only be done approximately, which strongly limits the usability of the resulting models for formal verification. 
\emph{Reachset conformance}, on the other hand, only requires that the system output is contained in the reachable set of the model at each time step, which facilitates using simpler model structures and smaller uncertainties.
As reachset conformance transfers safety properties {\cite{roehm2022reachset}}, a reachset-conformant model suffices for formal safety verification, e.g., for collision avoidance~\cite{althoff2019effortless} or the computation of control invariant sets~\cite{schaefer2024scalable}. 

Although reachset-conformant identification is a young research direction, there already exist multiple approaches to identify reachset-conformant state-space models in the literature: 
A given model can be made reachset-conformant by incrementally enlarging additive uncertainty sets until all measurements are contained in the reachable set~\cite{althoff2012reachability}.
If we have estimates for the system state, we can separately compute the additive process and measurement uncertainty set as multi-dimensional intervals that contain all errors between the system states and the states predicted by the model, and between the measurements and the model outputs~\cite{kochdumper2020conf}. 
Given full state measurements, the reachset-conformant identification of linear models with additive, zonotopic uncertainty sets can be formulated as a convex programming problem minimizing the size of the uncertainty sets~\cite{gruber2023scalable}.
Furthermore, additive, zonotopic uncertainty sets for linear systems can be computed with a linear program that minimizes the interval norm of the resulting reachable set~\cite{Liu2023conf}.
This method can be extended to the identification of the initial state uncertainty set and the minimization of the Frobenius norm of the reachable set~\cite{althoff2023conf}.
The scalability of the linear program with respect to the system dimension and time horizon can be improved by using the generator representation instead of the halfspace representation of the reachable set~\cite{luetzow2024generator}, {while recursive identification techniques improve the scalability with respect to the size of the identification dataset~\cite{luetzow2025recursive}.}

\subsection{Contributions and Outline}

To the best of our knowledge, there exist no publications about a) the reachset-conformant identification of arbitrary (non-additive) uncertainty sets, b) the reachset-conformant identification of input-output models such as autoregressive models with exogenous inputs (ARX models) or nonlinear autoregressive model with exogenous inputs (NARX models), and c) reachset-conformant black-box identification based entirely on data.
Our work addresses these topics by proposing a general framework for reachset-conformant system identification, as visualized in \cref{fig:id}. In particular, our contributions are as follows:
\begin{itemize}
    \item {We introduce a general linear output (GLO) approximation and derive its construction for state-space and input-output models.}
    \item We compute, for the first time, the reachable set of nonlinear input-output models using the {GLO approximation}.
    \item Furthermore, we present a linear program for the reachset-conformant white-box identification of linear and nonlinear models, i.e., we identify uncertainty sets such that reachset conformance is established. {To the best of our knowledge, this is the first approach to estimate, potentially} non-additive, uncertainty sets for input-output models and nonlinear state-space models.
    \item Additionally, we propose a framework for the identification of reachset-conformant models from data with little (gray-box identification) or no prior knowledge about the system dynamics (black-box identification).
    \item The performance and scalability of the proposed approaches are analyzed in numerical experiments for different dynamical systems and considering different levels of prior knowledge.
    \item We demonstrate real-world applicability by identifying a reachset-comformant vehicle model from real-world data.
\end{itemize}
This article is structured as follows:
Fundamentals about zonotopes and reachset conformance and the problem statement are introduced in \cref{sec:preliminaries}.
In \cref{sec:reach}, we present a general approach to compute the reachable set of different model types using the novel {GLO approximation}.
\cref{sec:idWhite} deals with the reachset-conformant identification of uncertainty sets for given model dynamics.
Subsequently, we extend this approach to cases where the model dynamics contain unknown parameters in \cref{sec:idGray} and cases where we do not have any knowledge about the model in \cref{sec:idBlack}.
The efficacy of our approaches is demonstrated in numerical experiments in \cref{sec:experiments}. \cref{sec:conclusions} concludes this article.

\subsection{Notation}
Sets are denoted by calligraphic letters, matrices by uppercase letters, and vectors and scalars by lowercase letters.
Furthermore, we use $\mathbf{1}$, $\mathbf{I}$, and $\mathbf{0}$, respectively, for a vector filled with ones, the identity matrix, and a matrix of zeros, where the matrix dimensions can be explicitly written in the subscript.
The operation $\mathrm{diag}(v)$ returns a diagonal matrix with the elements of the vector $v$ on its main diagonal and  $\mathrm{diag}(A_1, A_2, \dots, A_n)$ represents a blockdiagonal matrix where the matrices $A_1$, $A_2$, $\dots$, $A_n$ are concatenated diagonally. 
Additionally, we introduce the notation $v_{0:k}$ for the vertical concatenation $[v_0^\top~v_1^\top~\cdots~v_{k}^\top]^\top $ of the time-varying vector $v_i\in \mathbb{R}^m$ for the time steps $i=0,\dots,k$.
The Jacobian of $f(v_{0:k})\colon \mathbb{R}^{(k+1)m} \to \mathbb{R}^n$ with respect to the vector $v_i$ is written as 
$\nabla_{v_i}  f=[\frac{\partial f}{\partial v_{i,1}} ~ \cdots ~\frac{\partial f}{\partial v_{i,m}}]$, where $v_{i,j}$ denotes the $j$-th element of $v_i$. 

Let us also introduce the Minkowski sum of two sets $\mathcal{S}_a,~\mathcal{S}_b\subset \mathbb{R}^{n}$ as $\mathcal{S}_a \oplus \mathcal{S}_b = \{s_a + s_b | s_a \in \mathcal{S}_a,~s_b\in \mathcal{S}_b\}$, 
the Cartesian product $\mathcal{S}_a \times \mathcal{S}_b = \{ [s_a^\top ~ s_b^\top]^\top~|~ s_a \in \mathcal{S}_a, ~s_b \in \mathcal{S}_b \}$, and the linear transformation of $\mathcal{S}_a$ using the matrix $A \in \mathbb{R}^{m \times n}$ as $A\mathcal{S}_a = \{As | s \in \mathcal{S}_a\}$.
The Cartesian product $\mathcal{S}_0 \times \mathcal{S}_1 \times \dots \times \mathcal{S}_{k}$ of the sequence of sets $\mathcal{S}_i$, $i=0,\dots,k$, is denoted by $\mathcal{S}_{0:k}$.

\section{Preliminaries and Problem Statement}
\label{sec:preliminaries}

We first introduce zonotopes, which is the set representation used in this work, followed by the definition of reachset conformance. Subsequently, we formulate our problem statement.



\subsection{Zonotopes}

A zonotope is a convex set representation that is closed under Minkowski sum and linear map. The generator representation of zonotopes makes it possible to efficiently compute the aforementioned operations in high dimensions~\cite{girard2005zonotopes}:

\begin{definition}[Generator Representation of Zonotopes~\cite{kuehn1998wrapping}] \label{def:zon} 
A zonotope $\mathcal{Z}\subset \mathbb{R}^n$ can be described by its center vector $c =\mathrm{cen}(\mathcal{Z})\in \mathbb{R}^{n}$ and its generator matrix $G=\mathrm{gen}(\mathcal{Z})=[g_1~\cdots~g_{\eta}]\in \mathbb{R}^{n \times {\eta}}$ as
\begin{align*}
    \mathcal{Z} = \left\{c+\sum_{i=1}^{\eta} \lambda_i g_i \middle|\lambda_i\in[-1,1]\right\}=\langle  c,G \rangle.
\end{align*}
\end{definition}

Alternatively, zonotopes can be described by their halfspace representation:

\begin{definition}[Halfspace Representation of Zonotopes~\cite{Althoff2010halfspace}] \label{def:half}
A zonotope $\mathcal{Z}\subset \mathbb{R}^n$ can be represented as the intersection of a finite number of halfspaces, whose normal vectors are concatenated row-wise in the matrix $N\in \mathbb{R}^{h \times n}$ and whose offsets to the origin are stacked in the vector $d \in \mathbb{R}^{h}$:
\begin{align*}
\mathcal{Z} = \{x\in \mathbb{R}^{n} | Nx \leq d\}.
\end{align*}
\end{definition}

A zonotope in generator representation can be converted to its halfspace representation using~\cite[Thm.~7]{Althoff2010halfspace}.
Generally, computations using the generator representation scale better with increasing dimension than using the halfspace representation, since a zonotope with $\eta$ generators has up to $2\binom{\eta}{n-1}$ halfspaces~\cite{Althoff2010halfspace} and thus, the number of halfspaces $h$ grows exponentially with the dimension $n$. 
Furthermore, the set operations Minkowski sum, Cartesian product, and linear transformation can be efficiently performed in generator representation: 
For two zonotopes $\mathcal{Z}_a= \langle c_a,G_a\rangle$ and $\mathcal{Z}_b= \langle c_b,G_b\rangle$ and the matrix $A$, we have $\mathcal{Z}_a \oplus  \mathcal{Z}_b=\langle c_a+c_b,[G_a~G_b]\rangle$, $\mathcal{Z}_a \times  \mathcal{Z}_b=\langle [c_a^\top~c_b^\top]^\top,\mathrm{diag}(G_a, G_b)\rangle$, and $A\mathcal{Z}_a=\langle Ac_a,AG_a\rangle$.
The size of a zonotope can be evaluated using the interval norm:

\begin{definition}[Interval Norm of Zonotopes~{\cite[Sec.~3.1]{althoff2023conf}}]\label{def:intNorm} 
The interval norm for the zonotope $\mathcal{Z}=\langle c,G\rangle$ is defined as 
\begin{align*} 
\|\mathcal{Z}\|_I  = \mathbf{1}^\top |G| \mathbf{1}.
\end{align*}
\end{definition}


\subsection{Reachset Conformance} \label{subsec:reachConf}

As measurements are usually obtained in discrete time, we only consider discrete-time systems. 
Let us now introduce the output $\tau_k(\sys, x_{0}, u_{0:k})$ at time step $k$ of a system $\sys$ starting at the state $x_{0}$, controlled by the inputs stacked in $u_{0:k}= [u_0^\top ~\cdots~u_k^\top]^\top$. 
The input vector $u_i$ consists of the control signals and disturbances at time step $i$.
We consider uncertainties by assuming that the initial state $x_{0}$ and inputs $u_{i}$, $i=0,\dots,k$, deviate from the expected initial state $x_{*,0}$ and inputs $u_{*,i}$, which we call the nominal initial state and the nominal inputs. These deviations are represented by the sets $\uncX(\sys)$ and $\uncU{i}(\sys)$. The nominal initial state and inputs 
are collected in a test case:

\begin{definition}[Test Case] 
A test case {labeled by $m\in\mathcal{M}$} of length $n_{\mathrm{k}}$ consists of the nominal initial state $x_{*,0}^{(m)}$ and the nominal inputs $u_{*,0:n_{\mathrm{k}}-1}^{(m)}$.
\end{definition}

{Each test case is executed $n_{\mathrm{s}}$ times and the resulting outputs are measured as $y_k^{(m,s)}(\sys)$, with $k=0,...,n_{\mathrm{k}}-1$, $s=1,...,n_{\mathrm{s}}$.
While the number of executions $n_{\mathrm{s}}$ can vary for different test cases $m$, we assume it to be constant in this paper for notational simplicity.}

We refer to the set of reachable outputs as the reachable set, which can be formally defined as follows:
\begin{definition}[Reachable Set]\label{def:reach}
The reachable set of the system $\sys$ for test case $m$ and at time step $k$ is defined as
\begin{align*}
    {\Reach}_k^{(m)}(\sys) \coloneqq &\bigl\{\tau_k(\sys, x_{0}, u_{0:k}) \mid x_{0} \in x_{*,0}^{(m)}\oplus \uncX^{(m)}(\sys),\\   &\qquad  u_{0:k} \in u_{*,0:k}^{(m)}\oplus \uncU{0:k}^{(m)}(\sys) \bigr\}.
\end{align*}
\end{definition}

Subsequently, we consider a real-world system or a high-fidelity model as the target system $\sysT$. 
For many applications, such as real-time verification, we need a simple model $\sysM$ that is able to approximate the behavior of $\sysT$.
To transfer safety properties from the model $\sysM$ to the target system $\sysT$, reachset conformance is necessary and sufficient~\cite[Thm.~1]{roehm2022reachset}:
\begin{definition}[Reachset Conformance {\cite[Sec.~3.5]{roehm2019conformance}}]  \label{def:conf}
    System $\sysM$ is a reachset-conformant model of the target system $\sysT$ if and only if for all time steps $k$ and all possible  test cases $m$, the reachable set $\Reach_{k}^{(m)}(\sysM)$ contains the reachable set $\Reach_{k}^{(m)}(\sysT)$ of $\sysT$:
    \begin{align}
       &\sysT \, \mathtt{conf}_R \, \sysM \Leftrightarrow \notag\\
        \forall m, k\colon &
         \Reach_{k}^{(m)}(\sysT)  \subseteq  \Reach_{k}^{(m)}(\sysM). \label{eq:reachConf}
    \end{align}
\end{definition}
Since the reachable set does not exist for real-world systems and cannot be computed for many models \cite{Platzer2007,Gan2018reachability}, we use the following approximations: 
The reachable set of the target system $\sysT$ will be approximated by measurements $y^{(m,s)}_k(\sysT)$, i.e.,
\begin{align} 
    \Reach_k^{(m)}(\sysT) &= \lim_{n_{\mathrm{s}}\to \infty} \bigl\{y^{(m,s)}_k(\sysT)=\tau_k(\sysT, x_{0}^{(m,s)}, u_{0:k}^{(m,s)}) \,\big| \notag \\
    &\qquad s=1,\dots,n_{\mathrm{s}},~x_{0}^{(m,s)} \in x_{*,0}^{(m)}\oplus \uncX^{(m)}(\sysT), \notag \\
    &\qquad u_{0:k}^{(m,s)} \in u_{*,0:k}^{(m)}\oplus \uncU{0:k}^{(m)}(\sysT)\bigr\}. \label{eq:limy}
\end{align} 
As we can only consider a finite number of samples $n_{\mathrm{s}}$ in practice, reachset conformance can only be checked for the given samples but not be proven in general.
Furthermore, we might use the reachable set $\widehat{\Reach}_k^{(m)}(\sysM)$ computed by an overapproximative reachability algorithm instead of $\Reach_k^{(m)}(\sysM)$ in \cref{eq:reachConf}.
Safety properties can be transferred if the reachability algorithm used for verification is identical or even more overapproximative than the algorithm that was used for computing $\widehat{\Reach}_k^{(m)}(\sysM)$ when establishing reachset conformance. 


\subsection{Problem Statement}
We present methods for an efficient identification of reachset-conformant models from data.  
{A model $\sysM$ is characterized by a set of model functions $\func$, which describe the model dynamics and make it possible to predict the output $\tau_k(\sysM, x_{0}, u_{0:k})$ (see \cref{sec:reach} for definitions for different model types), and its uncertainty sets $\uncX^{(m)}(\sysM)$ and $\uncU{i}^{(m)}(\sysM)$. 
We assume that the uncertainty sets can be represented by test case- and time step-independent zonotopes. 
The center vectors consist of the initial estimates $c_{\mathrm{x}}\in \mathbb{R}^{n_{\mathrm{x}}}$ and $c_{\mathrm{u}}\in \mathbb{R}^{n_{\mathrm{u}}}$ and unknown center shifts $c_{\mathrm{\Delta},\mathrm{x}}\in \mathbb{R}^{n_{\mathrm{x}}}$ and $c_{\mathrm{\Delta},\mathrm{u}}\in \mathbb{R}^{n_{\mathrm{u}}}$. 
The generator matrices are constructed from known templates $G_{\mathrm{x}}\in \mathbb{R}^{n_{\mathrm{x}} \times \eta_{\mathrm{x}}}$ and $G_{\mathrm{u}} \in \mathbb{R}^{n_u \times \eta_{\mathrm{u}}}$, scaled by unknown factors $\alpha_{\mathrm{x}}\in \mathbb{R}^{\eta_x}_{\geq 0}$ and $\alpha_{\mathrm{u}}\in \mathbb{R}^{\eta_u}_{\geq 0}$. 
This results in the following uncertainty sets:}
\begin{subequations}\label{eq:uncSets}
\begin{align}
\uncX^{(m)}(\sysM) &= \langle c_{\mathrm{x}}+c_{\mathrm{\Delta},\mathrm{x}}, G_{\mathrm{x}}\mathrm{diag}(\alpha_{\mathrm{x}})\rangle,\\
\uncU{i}^{(m)}(\sysM) &= \langle c_{\mathrm{u}}+c_{\mathrm{\Delta},\mathrm{u}}, G_{\mathrm{u}} \mathrm{diag}(\alpha_{\mathrm{u}})\rangle,
\end{align}    
\end{subequations}
for all time steps $k=0,\dots,n_{\mathrm{k}}-1$ and test cases $m$. 
{For brevity, we introduce the combined variables $c=[c_{\mathrm{x}}^\top~ c_{\mathrm{u}}^\top]^\top$, $c_{\mathrm{\Delta}} = [c_{\mathrm{\Delta},\mathrm{x}}^\top~c_{\mathrm{\Delta},\mathrm{u}}^\top]^\top$, and $\alpha=[\alpha_{\mathrm{x}}^\top~\alpha_{\mathrm{u}}^\top]^\top$.}

We can now formulate our problem statement of identifying a reachset-conformant model.
Since unnecessary conservatism impedes the use of reachability analysis, we want the model to produce tight reachable sets. 
Thus, we additionally minimize a cost function that evaluates the size of the reachable sets for all test cases.

\begin{problem}[Reachset-conformant Identification]\label{prob:id}
\qquad 
\hphantom{WeWe}
We establish reachset conformance between the model $\sysMu{q}$ {with the uncertainty sets in \cref{eq:uncSets}} and the target system $\sysT$ for the test cases in $\mathcal{M}$. This is achieved by solving the optimization problem
\begin{subequations}
\begin{align}
q^* =& \argmin_{q} \sum_{m\in\mathcal{M}} \sum_{k=0}^{n_{\mathrm{k}}-1} w_k \, r\bigr(\widehat{\Reach}_{k}^{(m)}(\sysMu{q})\bigl)\label{eq:conf_cost}\\
\text{s.t. }&\quad \forall m, k,  s\colon y^{(m,s)}_k(\sysT) \in \widehat{\Reach}_{k}^{(m)}(\sysMu{q}), \label{eq:conf_constr}
\end{align}
\end{subequations}
where $q$ are unknown parameters of the uncertainty sets and the model functions of $\sysM$, 
the function $r(\mathcal{Y})$ evaluates the size of a set $\mathcal{Y}$, 
and $w_k\geq 0$ is a weight for time step $k$. 
\end{problem}

From this general problem, we derive the following subproblems considered in this work, which depend on the variables we optimize:

\begin{subproblem}[Identification of White-Box Models]\label{prob:idWhite}\qquad~~\hphantom{We}
We solve \cref{prob:id} with the optimization variables $q=(\alpha,~c_{\mathrm{\Delta}})$, assuming the {model functions $\func$} are known.
\end{subproblem}

\begin{subproblem}[Identification of Gray-Box Models]\label{prob:idGray}\qquad~~~~\hphantom{We}
We solve \cref{prob:id} with the optimization variables $q=(\alpha,~c_{\mathrm{\Delta}},~p)$, where the {model functions $\func=\func_p$} depend on unknown model parameters $p$.
\end{subproblem}

\begin{subproblem}[Identification of Black-Box Models]\label{prob:idBlack}\qquad\hphantom{We}
We solve \cref{prob:id} with the optimization variables $q=(\alpha,~c_{\mathrm{\Delta}},~\func)$. 
\end{subproblem}

Estimating arbitrary model parameters $p$ in \cref{prob:idGray} or the {functions $\func$} in \cref{prob:idBlack} are complex, nonlinear programming problems. 
{As computing the global optimal solution of general nonlinear programming problems is NP-hard and hence intractable, we will focus on efficiently finding a good local optimum that satisfies the conformance constraints.

Note that we will omit the notation $(\sysM)$ for variables referring to the model $\sysM$ in the remainder of this work to enhance readability.}

\section{Reachability Analysis}\label{sec:reach}

{In this section, we explain how to generalize reachability analysis for different model types using a so-called general linear output approximation.
\begin{definition}[{General Linear Output (GLO) Approximation}] \label{def:linearizedModel} 
    {The GLO approximation }
    \begin{align}
        {\bar{\tau}_k^{(m)}(x_{0}, u_{0:k})} &= \bar{y}_k^{(m)} + \bar{C}_k^{(m)} (x_{0}-\bar{x}_{0}^{(m)}) \notag \\&\quad + \sum_{i=0}^k \bar{D}_{k,i}^{(m)} (u_{i}-\bar{u}_{i}^{(m)})\label{eq:GLOmodel}
    \end{align} 
    \color{blue}is defined by the reference output $\bar{y}_k^{(m)}$, the system matrix $\bar{C}_k^{(m)}$, and the input matrices $\bar{D}_{k,i}^{(m)}$. 
    It provides a linear approximation of the output $\tau_k$ along the reference trajectory defined by $\bar{x}_{0}^{(m)}$ and $\bar{u}_{0:k}^{(m)}$.
    The deviation between the output $\tau_k$ and the linearized output $\bar{\tau}_k^{(m)}$ is described by the linearization error $\bar{e}_k^{(m)}$:
    \begin{align*}
        \bar{e}_k^{(m)}(x_{0}, u_{0:k})=\tau_k(x_{0}, u_{0:k})-\bar{\tau}_k^{(m)}(x_{0}, u_{0:k}).
    \end{align*}
\end{definition}

Using the GLO approximation, we can simplify reachability analysis as follows:
\begin{theorem}[Reachability using the GLO approximation] \label{theo:reach} 
    The reachable set of outputs $\tau_k$ for test case $m$ can be overapproximated 
    as 
    \begin{align}
    	\widehat{\Reach}_k^{(m)} = \bar{\Reach}_k^{(m)} \oplus \widehat{\mathcal{E}}_k^{(m)}
    	\label{eq:reachability},
    \end{align}
    where  
    \begin{align}
        \bar{\Reach}_k^{(m)} &= \bar{y}_k^{(m)} \oplus \bar{C}_k^{(m)} (-c_{\mathrm{x}}\oplus \uncX^{(m)}) \notag\\&\qquad \oplus \bigoplus_{i=0}^k \bar{D}_{k,i}^{(m)} (-c_{\mathrm{u}}\oplus \uncU{i}^{(m)}), \label{eq:reachability_GLO}
    \end{align}
    and $\bar{y}_k^{(m)}$, $\bar{C}_k^{(m)}$, and $\bar{D}_{k,i}^{(m)}$ are the reference output, system matrix, and input matrices of the GLO approximation of $\tau_k$ along the reference trajectory defined by      
    \begin{subequations} \label{eq:refTraj}    
        \begin{align}
    	\bar{x}_{0}^{(m)} &= x_{*,0}^{(m)}+ c_{\mathrm{x}},\label{eq:refTraj_x}  \\
    	\bar{u}_i^{(m)} &=u_{*,i}^{(m)}+ c_{\mathrm{u}}, ~i=0,\dots,k.\label{eq:refTraj_u} 
        \end{align}%
    \end{subequations}
    The vectors $c_{\mathrm{x}}$ and $c_{\mathrm{u}}$ are the initial estimates for the center vectors of the uncertainty sets $\uncX^{(m)}$ and $\uncU{i}^{(m)}$, and the set $\widehat{\mathcal{E}}_k^{(m)}$ is an enclosure of the linearization errors $\bar{e}_k^{(m)}$. 
    The overall tightness of the overapproximation $\widehat{\Reach}_k^{(m)}\supseteq {\Reach}_k^{(m)}$ depends solely on the conservatism introduced by $\widehat{\mathcal{E}}_k^{(m)}$.

\begin{proof}
    Using \cref{def:linearizedModel}, the output for test case $m$ can be expressed as
    \begin{align}\label{eq:tau_e}
    	\tau_k(x_{0}, u_{0:k}) = \bar{\tau}_k^{(m)}( x_{0}, u_{0:k}) + \bar{e}_k^{(m)}(x_{0}, u_{0:k}).
    \end{align} 
    A set-based evaluation of \cref{eq:tau_e} over the uncertain initial state $x_0 \in x_{*,0}\oplus \uncX^{(m)}$ and the uncertain inputs $u_{i}\in u_{*,i}^{(m)}\oplus \uncU{i}^{(m)}$, $i=0,\dots,k$, leads to \cref{eq:reachability}, where the reachable set of the GLO approximation 
    \begin{align*}
        \bigl\{\bar{\tau}_k^{(m)}( x_{0}, u_{0:k}) \mid x_{0} \in x_{*,0}^{(m)}\oplus \uncX^{(m)},  u_{0:k} \in u_{*,0:k}^{(m)}\oplus \uncU{0:k}^{(m)} \bigr\}
    \end{align*}
    is denoted by $\bar{\Reach}_k^{(m)}$, and the set $\widehat{\mathcal{E}}_k^{(m)}$ encloses all linearization errors.
    Eq.~\cref{eq:reachability_GLO} can be obtained from a set-based evaluation of~\cref{eq:GLOmodel} and the reference trajectory in \eqref{eq:refTraj} centers the linearization at the expected midpoint of the initial state and input sets to reduce the expected linearization error.

    Although the center shifts $c_{\mathrm{\Delta},\mathrm{x}}$ and $c_{\mathrm{\Delta},\mathrm{u}}$ are unknown, they are treated as point values. This modeling choice implies that dependencies between uncertainties at different time steps are not considered (i.e., each uncertain variable takes values from its respective uncertainty set independently of the others), thereby avoiding any additional overapproximation that could arise from ignoring such dependencies \cite{finkbeiner2022monitoring}.
    Moreover, since the uncertain initial state and all uncertain input variables appear only once in the GLO approximation, we do not encounter the dependency effect \cite[Sec.~2.2.3]{jaulin2001interval} in \cref{eq:reachability_GLO}, which would otherwise introduce additional conservatism. 
    As a result, \cref{eq:reachability_GLO} computes the exact reachable set of the GLO approximation, and the overall tightness of $\widehat{\Reach}_k^{(m)}$ depends solely on the conservatism introduced by $\widehat{\mathcal{E}}_k^{(m)}$.
\end{proof}
\end{theorem}
 
Using this theorem, we can compute the reachable set for different model types, where we compute the GLO approximation $\bar{\tau}_k^{(m)}$ from a first-order Taylor series expansion of the output $\tau_k$ along the reference trajectory to obtain small linearization errors.
In the following subsections, we present the formulas for computing the GLO approximation and the set of linearization errors for state-space and input-output models.}
For improved readability, we omit the test case $m$ in the superscripts in the following derivations.

\subsection{State-Space Models}
A popular model type, which is especially used in control applications, is the state-space model. In the general nonlinear case, {it is described by the model functions $\func=\{f,g\}$}, where
\begin{subequations} \label{eq:nss}
	\begin{align}
		x_{k+1} &= f(x_k, u_k) \label{eq:nss_x}\\
		y_k &= g(x_k,u_k).\label{eq:nss_y}
	\end{align}    
\end{subequations}

\begin{corollary}[Reachability of State-Space Models] \label{cor:conv_ss}
    The reachable set of the state-space model in~\cref{eq:nss} can be computed using \cref{theo:reach}, with  
    \begin{subequations}\label{eq:ss_yCD}        
    \begin{align}
        \bar{y}_k &= g(\bar{x}_k,\bar{u}_k), \label{eq:ss_yref} \\
        \bar{C}_k &= C_{k} \prod_{l=1}^{k} A_{k-l}, \label{eq:ss_C}\\
        \bar{D}_{k,i} &= \begin{cases}
            C_{k}\left(\prod_{l=1}^{k-i-1} A_{k-l}\right) B_{i} & 0\leq i < k, \\
            D_{k} & i=k,
            \end{cases}\label{eq:ss_D}\\  
            \widehat{\mathcal{E}}_k &= C_{k}\bigoplus_{j=0}^{k-1}
		\left(\prod_{l=1}^{k-j-1} A_{k-l}\right)\widehat{\mathcal{L}}_{\mathrm{x},j}  \oplus \widehat{\mathcal{L}}_{\mathrm{y},k}. \notag
    \end{align}
    \end{subequations}
    The reference output in \cref{eq:ss_yref} is obtained from simulating \cref{eq:nss} along the reference trajectory in \cref{eq:refTraj},
    the sets $\widehat{\mathcal{L}}_{\mathrm{x},j}$ and $\widehat{\mathcal{L}}_{\mathrm{y},k}$ can be computed as in \cite[Sec.~3.1]{althoff2013nonlin}, and
	\begin{align*}
		A_{k} &= \nabla_{{x}_k}~f(x_k,u_k)|_{\bar{x}_{k}, \bar{u}_k}, ~~
		B_{k} =  \nabla_{{u}_k}~f(x_k,u_k)|_{\bar{x}_{k}, \bar{u}_k}, \\
		C_{k} &= \nabla_{{x}_k}~g(x_k,u_k)|_{\bar{x}_{k}, \bar{u}_k}, ~~
		D_{k} =  \nabla_{{u}_k}~g(x_k,u_k)|_{\bar{x}_{k}, \bar{u}_k}.
	\end{align*} 
 
	\begin{proof}
		Applying a Taylor expansion of order $\kappa$ with Lagrange remainder to \cref{eq:nss} along the reference trajectory, we obtain
		\begin{align}
			x_{k+1}
			&= f(\bar{x}_k, \bar{u}_k) + A_{k} (x_k - \bar{x}_k)  + B_{k} (u_k - \bar{u}_k) + {l}_{\mathrm{x},k} \label{eq:sys_lin_x}\\
			y_k
			&= g(\bar{x}_k,\bar{u}_k) + C_{k} (x_k - \bar{x}_k)  + D_{k} (u_k - \bar{u}_k) + {l}_{\mathrm{y},k}, \label{eq:sys_lin_y}
		\end{align}
		where $l_{\mathrm{x},k}\in\mathcal{L}_{\mathrm{x},k}$ and $l_{\mathrm{y},k}\in\mathcal{L}_{\mathrm{y},k}$ contain the Taylor terms from order two up to order $\kappa$ and the Lagrange remainders~\cite{althoff2013nonlin}.
		Recursively inserting \cref{eq:sys_lin_x} in \cref{eq:sys_lin_y} leads to the {GLO approximation} defined by the formulas above with
		\begin{align*}
			\bar{e}_k = C_{k}\sum_{j=0}^{k-1}
			\Biggl(\prod_{l=1}^{k-j-1} A_{k-l}\Biggr)l_{\mathrm{x},j} + l_{\mathrm{y},k}.
		\end{align*}
		The set $\widehat{\mathcal{E}}_k$ enclosing the linearization errors is computed by a set-based evaluation of $\bar{e}_k$ using the overapproximations $\widehat{\mathcal{L}}_{\mathrm{x},j}\supseteq \mathcal{L}_{\mathrm{x},j}$ and $\widehat{\mathcal{L}}_{\mathrm{y},k}\supseteq \mathcal{L}_{\mathrm{y},k}$ from \cite[Sec.~3.1]{althoff2013nonlin}.
	\end{proof}
\end{corollary}
If the state-space model is linear, we have $\widehat{\mathcal{E}}_k=\mathbf{0}$.

\subsection{Input-Output Models}
To compute the reachable set of state-space models, we require some estimate for the initial state $x_{*,0}$, which is often not directly measurable. 
In that case, guaranteed state estimation \cite{Althoff2021comp} is required, which introduces additional uncertainties. 
{By contrast, input–output models can be directly initialized from measurements, avoiding the need for state estimation and often resulting in tighter reachable sets \cite{luetzow2023armax}. 
This makes input–output models a valuable alternative for data-driven system identification.

A widely used class of input–output models is the nonlinear autoregressive model with exogenous inputs (NARX model), where the current output is computed from the current input and a finite history of past outputs and inputs using the model function $\func=\{h\}$:
\begin{align}
	y_k &= h(y_{k-{n_{\mathrm{p}}}:k-1}, u_{k-{n_{\mathrm{p}}}:k}), \label{eq:narx}
\end{align}
and $n_{\mathrm{p}}$ denotes the number of past time steps considered. 
The model can be recursively evaluated from initial measurements, which form the initial state vector 
		$x_0= \begin{bmatrix}
			y_0^\top &
			y_1^\top &
			\cdots &
			y_{n_{\mathrm{p}}-1}^\top 
		\end{bmatrix}^\top$.
}
By combining a Taylor expansion with the reformulation from \cite[Thm.~1]{luetzow2023armax}, we obtain the following formulas to compute the reachable set:

\begin{corollary}[Reachability of Input-Output Models] \label{cor:conv_io} 
    The reachable set of the input-output model in~\cref{eq:narx} can be computed using \cref{theo:reach}, with $k\geq{n_{\mathrm{p}}}$ and
    \begin{subequations}\label{eq:io_yCD}        
    \begin{align}
        \bar{y}_k &= h(\bar{y}_{k-{n_{\mathrm{p}}}:k-1}, \bar{u}_{k-{n_{\mathrm{p}}}:k}), \label{eq:io_yref}\\
        \bar{C}_k &= E \prod_{l=0}^{k-{n_{\mathrm{p}}}} {A}_{\mathrm{ext},k-l}, \label{eq:io_C}\\
        \bar{D}_{k,i} &= E \sum_{j=0}^{k-{n_{\mathrm{p}}}} \Biggl( \prod_{l=0}^{j-1} {A}_{\mathrm{ext},k-l}\Biggr) {B}_{\mathrm{ext},k-j,k-i-j},\label{eq:io_D}\\  
        \widehat{\mathcal{E}}_k &= E \bigoplus_{j=0}^{k-{n_{\mathrm{p}}}} \Biggl( \prod_{l=0}^{j-1} {A}_{\mathrm{ext},k-l}\Biggr) E^\top \widehat{\mathcal{L}}_{k-j}. \notag
    \end{align}    
    \end{subequations}
    The reference output in \cref{eq:io_yref} is obtained by simulating \cref{eq:narx} along the reference trajectory in \cref{eq:refTraj}, the set $\widehat{\mathcal{L}}_{k}$ can be computed analogously to \cite[Sec.~3.1]{althoff2013nonlin}, and
    \begin{align*}  
        E &= \begin{bmatrix}
		\mathbf{0}_{n_{\mathrm{y}}\times ({n_{\mathrm{p}}}-1)n_{\mathrm{y}}} &
		\mathbf{I}_{n_{\mathrm{y}}}   
		\end{bmatrix},\\
        {A}_{\mathrm{ext},k} &=\begin{bmatrix}  
            \mathbf{0}_{({n_{\mathrm{p}}}-1)n_{\mathrm{y}}\times n_{\mathrm{y}}} &\mathbf{I}_{({n_{\mathrm{p}}}-1)n_{\mathrm{y}}}  \\
            {A}_{k,{n_{\mathrm{p}}}} & [{A}_{k,{n_{\mathrm{p}}}-1}~ \cdots~{A}_{k,1}]
            \end{bmatrix}, \\
        {B}_{\mathrm{ext},k,i} &= \begin{bmatrix}
            \mathbf{0}_{({n_{\mathrm{p}}}-1)n_{\mathrm{y}}\times n_u}       \\
            {B}_{k,i}
            \end{bmatrix},\\
	A_{k,i} &= \nabla_{y_{k-i}}~h(y_{k-{n_{\mathrm{p}}}:k-1}, u_{k-{n_{\mathrm{p}}}:k})|_{\bar{y}_{k-{n_{\mathrm{p}}}:k-1},\bar{u}_{k-{n_{\mathrm{p}}}:k}}, \\
        {B}_{k,i} &= \nabla_{{u}_{k-i}}~h(y_{k-{n_{\mathrm{p}}}:k-1}, u_{k-{n_{\mathrm{p}}}:k})|_{\bar{y}_{k-{n_{\mathrm{p}}}:k-1},\bar{u}_{k-{n_{\mathrm{p}}}:k}}.
    \end{align*}

	\begin{proof}
		The output of the NARX model can be approximated by a Taylor series of order $\kappa$ along the reference trajectory defined by $\bar{x}_0= [\bar{y}_0^\top~\bar{y}_1^\top~\cdots~
	\bar{y}_{n_{\mathrm{p}}-1}^\top]^\top$ and $\bar{u}_{0:k}$:
		\begin{align*}
			y_k &= h(\bar{y}_{k-{n_{\mathrm{p}}}:k-1}, \bar{u}_{k-{n_{\mathrm{p}}}:k}) + \sum_{i=1}^{n_{\mathrm{p}}} A_{k,i} (y_{k-i} - \bar{y}_{k-i}) \\
			&\quad +\sum_{i=0}^{n_{\mathrm{p}}} B_{k,i} (u_{k-i} - \bar{u}_{k-i}) + {l}_{k}, 
		\end{align*}
		where $l_{k}\in \mathcal{L}_{k}$ contains the Taylor terms from order two up to order $\kappa$ and the Lagrange remainder~\cite{althoff2013nonlin}.
		A reformulation analog to \cite[Thm.~1]{luetzow2023armax} results in the {GLO approximation} defined by the formulas above with 
	\begin{align*}
			\bar{e}_k = E \sum_{j=0}^{k-{n_{\mathrm{p}}}} \Biggl(\prod_{l=0}^{j-1} {A}_{\mathrm{ext},k-l}\Biggr) E^\top {l}_{k-j}.
	\end{align*}
	A set-based evaluation of $\bar{e}_k$ leads to the set $\widehat{\mathcal{E}}_k$ using the overapproximation $\widehat{\mathcal{L}}_{k}\supseteq \mathcal{L}_{k}$ from \cite[Sec.~3.1]{althoff2013nonlin}.
	\end{proof}
\end{corollary}
If the function {$h$} in \cref{eq:narx} is linear in $y_{k-{n_{\mathrm{p}}}:k-1}$ and $u_{k-{n_{\mathrm{p}}}:k}$, i.e., the model is an ARX model, we have $\widehat{\mathcal{E}}_k=\mathbf{0}$.


\section{Identification of White-Box Models}
\label{sec:idWhite}
This section describes the general framework to identify {uncertainty sets that ensure reachset conformance for models with known model functions $\func$} (\cref{prob:idWhite}).
For linear models and the zonotopic uncertainty sets in \cref{eq:uncSets}, reachset conformance can be established by computing scaling factors $\alpha$ and center shifts $c_{\mathrm{\Delta}}$ with linear programming~\cite{luetzow2024generator}.
This approach can be generalized to a nonlinear model by computing its reachable set using \cref{theo:reach} for each test case $m$. 
To formulate the white-box identification problem from \cref{prob:idWhite} as a linear program (LP), we require the following lemmas.
\begin{lemma}[Set of Output Deviations]\label{lem:Ya}
    Given the uncertainty sets in \cref{eq:uncSets}, 
    the center vector and generators of the set $\bar{\Reach}_{k}^{(m)}$ from \cref{theo:reach} are
    \begin{subequations} \label{eq:cG_Ya}       
    \begin{align}
        \mathrm{cen}\bigl(\bar{\Reach}_{k}^{(m)}\bigr) &= \bar{y}_k^{(m)} + \bar{C}_k^{(m)} c_{\mathrm{\Delta},\mathrm{x}}+\sum_{i=0}^k \bar{D}_{k,i}^{(m)}c_{\mathrm{\Delta},\mathrm{u}},\label{eq:c_Ya}\\
        \mathrm{gen}\bigl(\bar{\Reach}_{k}^{(m)}\bigr) 
        &= \mathrm{gen}'\bigl(\bar{\Reach}_{k}^{(m)}\bigr) \mathrm{diag}\Bigl(\begin{bmatrix}
            \alpha_{\mathrm{x}}^\top &\alpha_{\mathrm{u}}^\top &\cdots&\alpha_{\mathrm{u}}^\top
            \end{bmatrix}^\top \Bigr), \label{eq:G_Ya}\\  
        \mathrm{gen}'\bigl(\bar{\Reach}_{k}^{(m)}\bigr)&=\begin{bmatrix}
            \bar{C}_k^{(m)} G_{\mathrm{x}} & \bar{D}_{k,0}^{(m)} G_{\mathrm{u}}&\cdots&\bar{D}_{k,k}^{(m)} G_{\mathrm{u}}
        \end{bmatrix}. \label{eq:Ghat_Ya}
    \end{align}
    \end{subequations}
    
    \begin{proof}    
        We insert the uncertainty sets from \cref{eq:uncSets} 
        in \cref{eq:reachability_GLO}.   
    \end{proof}
\end{lemma}
The cost function in \cref{eq:conf_cost} can be expressed linearly in $\alpha$ as follows:
\begin{lemma}[Cost Function]\label{lem:cost}
    Given \cref{lem:Ya} and using $r\bigr(\widehat{\Reach}_{k}^{(m)}\bigl)=\|\bar{\Reach}_{k}^{(m)}\|_I$, the cost in \cref{eq:conf_cost} can be computed as $\gamma \alpha$, where $\alpha \geq \mathbf{0}$ and
    \begin{align}
		\gamma &= \sum_{m\in\mathcal{M}}\sum_{k=0}^{n_{\mathrm{k}}-1} w_k \mathbf{1}^\top \Bigl[
		|\bar{C}^{(m)}_k G_{\mathrm{x}}|~~ \sum_{i=0}^{k}|\bar{D}_{k,i}^{(m)}G_{\mathrm{u}} |
		\Bigr]. \label{eq:gamma}  
    \end{align}    
    \begin{proof}
        We evaluate \cref{eq:conf_cost}, where $\|\bar{\Reach}_{k}^{(m)}\|_I$ is computed by inserting \cref{eq:G_Ya} into \cref{def:intNorm}, analogously to \cite[Lem.~1]{luetzow2024generator}.
    \end{proof}
\end{lemma}

To enforce the conformance constraints in \cref{eq:conf_constr}, we use one of the following lemmas, which are based on the halfspace representation or the generator representation of $\bar{\Reach}_{k}^{(m)}$. 
\begin{lemma}
[Halfspace Constraints] \label{lem:half} 
Given \cref{lem:Ya}, the containment constraint $y^{(m,s)}_k(\sysT) \in \bar{\Reach}_{k}^{(m)}$, $\forall s$, is equivalent to
\begin{align}
 \max_{s}\left(N^{(m)}_k y_{\mathrm{a},k}^{(m,s)}\right)  &\leq P_{\mathrm{\alpha},k}^{(m)} \alpha + P_{\mathrm{c},k}^{(m)} c_{\mathrm{\Delta}}, \label{eq:conf_constr_half1} 
\end{align}
with $\alpha \geq \mathbf{0}$, $y_{\mathrm{a},k}^{(m,s)}= y_{k}^{(m,s)}(\sysT)-  \bar{y}_k^{(m)}$, and
\begin{align}    
    P_{\mathrm{\alpha},k}^{(m)} &= \Bigl[
     | N^{(m)}_k \bar{C}^{(m)}_k G_{\mathrm{x}} | ~~ \sum_{i=0}^{k}  |N^{(m)}_k\bar{D}_{k,i}^{(m)}G_{\mathrm{u}} |\Bigr] \notag, \\
    P_{\mathrm{c},k}^{(m)} &= \Bigl[ N^{(m)}_k \bar{C}^{(m)}_k ~~ N^{(m)}_k \sum_{i=0}^{k} \bar{D}_{k,i}^{(m)}
    \Bigr].   \notag
\end{align}
The rows of $N^{(m)}_k$ are the normal vectors of the halfspace representation of $\bar{\Reach}_{k}^{(m)}$, which can be obtained using~\cite[Thm.~7]{Althoff2010halfspace}.

\begin{proof}   
By subtracting $\bar{y}_k^{(m)}$ on both sides, we can write the constraint $y_{k}^{(m,s)}(\sysT)\in  \bar{\Reach}_{k}^{(m)}$ as $y_{\mathrm{a},k}^{(m,s)}\in -\bar{y}_k^{(m)} \oplus\bar{\Reach}_{k}^{(m)}$.
With the halfspace representation of $\bar{\Reach}_{k}^{(m)}$, where $\bar{\Reach}_{k}^{(m)}$ can be computed using \cref{lem:Ya}, this constraint can be enforced for each test case $m$ analogously to \cite[Thm.~1]{luetzow2024generator}. 
\end{proof}
\end{lemma}

\begin{lemma}[Generator Constraints]\label{lem:gen}
Given \cref{lem:Ya}, the containment constraint $y^{(m,s)}_k(\sysT) \in \bar{\Reach}_{k}^{(m)}$, $\forall s$, is equivalent to
\begin{subequations}\label{eq:conf_opt_gen}
\begin{align}
    \forall s\colon \quad 
  \begin{bmatrix}\mathbf{0} \\ \mathbf{0} \end{bmatrix} &\leq \begin{bmatrix}
            R_{\mathrm{\alpha},k} \alpha + \beta^{(m,s)}_{k} \\
            R_{\mathrm{\alpha},k} \alpha - \beta^{(m,s)}_{k} \\
        \end{bmatrix}, \label{eq:conf_constr_gen1}\\
    y_{\mathrm{a},k}^{(m,s)} &= Q_{\mathrm{c},k}^{(m)}c_{\mathrm{\Delta}} + Q_{\mathrm{\beta},k}^{(m)} \beta^{(m,s)}_{k} \label{eq:conf_constr_genEq},
\end{align}
\end{subequations}
with $\alpha \geq \mathbf{0}$, $y_{\mathrm{a},k}^{(m,s)}= y_{k}^{(m,s)}(\sysT)-  \bar{y}_k^{(m)}$,  and the auxiliary optimization variables $\beta^{(m,s)}_{k}\in\mathbb{R}^{\eta_x+(k+1)\eta_u}$.
The constraint matrices are
\begin{subequations}  \label{eq:R_alpha}   
\begin{align*}
Q_{\mathrm{c},k}^{(m)} &= \begin{bmatrix}
    \bar{C}_{k}^{(m)} &\sum_{i=0}^k \bar{D}_{k,i}^{(m)}\end{bmatrix}, \notag \\
Q_{\mathrm{\beta},k}^{(m)} &= \mathrm{gen}'\bigl({\bar{\Reach}_{k}^{(m)}}\bigr), \notag\\
R_{\mathrm{\alpha},k} &=\begin{bmatrix}
    \mathbf{I}_{\eta_{\mathrm{x}}} & \mathbf{0}_{\eta_{\mathrm{x}} \times \eta_{\mathrm{u}}} \\
    \mathbf{0}_{(k+1)\eta_{\mathrm{u}} \times \eta_{\mathrm{x}}} & \mathrm{vert}_{k+1}\left( \mathbf{I}_{\eta_{\mathrm{u}}}\right)
\end{bmatrix},
\end{align*}
\end{subequations}
where $\mathrm{vert}_{k+1}\left( \mathbf{I}_{\eta_{\mathrm{u}}}\right)$ is the vertical concatenation of $k+1$ identity matrices and $\mathrm{gen}'\bigl({\bar{\Reach}_{k}^{(m)}}\bigr)$ is computed as in \cref{eq:Ghat_Ya}.

\begin{proof}
   We can write the constraint $y_{k}^{(m,s)}(\sysT)\in  \bar{\Reach}_{k}^{(m)}$ as $y_{\mathrm{a},k}^{(m,s)}\in -\bar{y}_k^{(m)}\oplus \bar{\Reach}_{k}^{(m)}$ after subtracting $\bar{y}_k^{(m)}$ on both sides. This constraint can be enforced via the generator representation of $\bar{\Reach}_{k}^{(m)}$ analogously to \cite[Thm.~2]{luetzow2024generator}, i.e.,
    \begin{subequations}        \label{eq:beta}
    \begin{align}
        y_{\mathrm{a},k}^{(m,s)}\in& -\bar{y}_k^{(m)}\oplus \bar{\Reach}_{k}^{(m)} \notag \\
        \overset{\text{\cref{def:zon}}}{\Leftrightarrow}  ~&\exists \lambda^{(m,s)}_{k} \in \mathbb{R}^{\eta_{\mathrm{x}}+(k+1) \eta_{\mathrm{u}}}\colon \notag \\ 
        &\quad |\lambda^{(m,s)}_{k}| \leq \textbf{1},  \notag\\ 
        &\quad
    y_{\mathrm{a},k}^{(m,s)} = -\bar{y}_k^{(m)}+\mathrm{cen}\bigl({\bar{\Reach}_{k}^{(m)}}\bigr)+ \mathrm{gen}\bigl({\bar{\Reach}_{k}^{(m)}}\bigr)\lambda^{(m,s)}_{k} \notag     \\
        \overset{~\text{\cref{eq:cG_Ya}}~}{\Leftrightarrow}~& \exists \beta^{(m,s)}_{k}\in \mathbb{R}^{\eta_{\mathrm{x}}+(k+1) \eta_{\mathrm{u}}}\colon \notag \\ 
            &\quad
        |\beta^{(m,s)}_{k}| \leq \begin{bmatrix}
                \alpha_{\mathrm{x}}^\top &\alpha_{\mathrm{u}}^\top &\cdots&\alpha_{\mathrm{u}}^\top
            \end{bmatrix}^\top, \label{eq:beta1}\\ 
            &\quad
        y_{\mathrm{a},k}^{(m,s)} = 
                \bar{C}^{(m)}_kc_{\mathrm{\Delta},\mathrm{x}} +\sum_{i=0}^k \bar{D}_{k,i}^{(m)}c_{\mathrm{\Delta},\mathrm{u}} \notag \\
                & \qquad \qquad + \mathrm{gen}'\bigl({\bar{\Reach}_{k}^{(m)}}\bigr) \beta^{(m,s)}_{k}, \label{eq:beta2}
    \end{align}    
    \end{subequations}
    where we define $\beta^{(m,s)}_{k}\coloneqq\mathrm{diag}([\alpha_{\mathrm{x}}^\top~~\alpha_{\mathrm{u}}^\top ~~\cdots~~ \alpha_{\mathrm{u}}^\top]^\top) \lambda^{(m,s)}_{k}$. 
    Finally, the constraints in~\cref{eq:beta1,eq:beta2} can be written as in~\cref{eq:conf_constr_gen1} and~\cref{eq:conf_constr_genEq}.
\end{proof}
\end{lemma}

{The resulting white-box identification procedure is summarized in \cref{alg:white}. It takes as input the set of model functions $\func$ (i.e., $\func=\{f,g\}$ from \cref{eq:nss} for state-space models or $\func=\{h\}$ from \cref{eq:narx} for input-output models), the matrix $B_c$ that determines which components of the center shift $c_{\mathrm{\Delta}}$ are to be identified, and the set of test cases $\mathcal{M}$.
In \cref{line:CDcomp}, we compute the GLO approximations $\bar{\tau}^{(m)}$ for all $m\in\mathcal{M}$ using the function \texttt{computeGLO()}, which implements \cref{eq:ss_yCD} for state-space models or \cref{eq:io_yCD} for input-output models. 
The scaling factors $\alpha$ and center shifts $c_{\mathrm{\Delta}}$ are then identified by solving the linear program in \cref{eq:conf_opt}, as shown in \cref{line:whiteLP}.
This optimization minimizes the cost function defined in \cref{lem:cost}, subject to the reachset-conformance constraints provided by} {\cref{lem:half} or \cref{lem:gen}.
The matrix $B_c$, a diagonal matrix with entries in $\{0,1\}$, encodes which elements of $c_{\mathrm{\Delta}}$ are fixed to zero via the constraint in \cref{eq:deltaCconstraint}. 
Since estimating optimal center shifts for nonlinear models would require introducing nonlinear constraints (see the proof of \cref{theo:White}), we enforce $c_{\mathrm{\Delta}}=\mathbf{0}$ for such models by setting $B_c$ to the identity matrix in \cref{line:whiteBc1,line:whiteBc2,line:whiteBcEnd}.}
To identify the center vectors for nonlinear models, we can consider $c$ as a part of the unknown model parameters $p$ and use the methods presented in the next section.

\begin{algorithm}[bt]
\caption{\texttt{White}.}\label{alg:white} 
\begin{algorithmic}[1]
\Statex \textbf{Inputs:}
  {\Desc{~~$\func$}{Model functions}  
  \Desc{~~$B_c$}{Center identification matrix}} 
  \Desc{~~$\mathcal{M}$}{Test cases}  
\Statex \textbf{Outputs:}
  \Desc{~~$\alpha^*$}{Optimized scaling factors}
  \Desc{~~$c_{\mathrm{\Delta}}^*$}{Optimized center shifts}
  \Desc{~~$\ell_{\mathrm{C}}^*$}{Final conformance cost}  
\vspace{0.2cm}
\Statex \comm{Compute GLO approximation for each test case}
\State {$\bar{\tau}^{(m)} \gets$ \texttt{computeGLO}$(\func, m), \forall m\in\mathcal{M}$}\label{line:CDcomp}
\Statex \comm{Optimize scaling factors and center shifts}
\State $\gamma \gets$ \cref{eq:gamma} 
\Statex \comm{Ensure $B_{\mathrm{c}}$ is identity matrix for nonlinear model functions}
\If{{any function in $\func$ is nonlinear}}\label{line:whiteBc1}
    \State $B_{\mathrm{c}} \gets \mathbf{I}$\label{line:whiteBc2}
\EndIf\label{line:whiteBcEnd}
\State $\alpha^*,c_{\mathrm{\Delta}}^* \gets$ \label{line:whiteLP}\vspace{-0.65cm}
    \begin{subequations}\label{eq:conf_opt}
	\begin{align}
			\hspace{1.2cm}& \argmin_{\alpha, c_{\mathrm{\Delta}}, (\beta)} \gamma \alpha && \qquad \label{eq:confSimpl_cost}\\
            \text{s.t. } &\quad  \text{constraint } \cref{eq:conf_constr_half1} \text{ or } \cref{eq:conf_opt_gen},~\forall m,k, \label{eq:contConstraint}\\
            &\quad \alpha \geq \mathbf{0} 	\label{eq:posconstraint},	\\
            &\quad  B_{\mathrm{c}} c_{\mathrm{\Delta}} = \mathbf{0} \label{eq:deltaCconstraint}, 
	\end{align}
    \end{subequations} \vspace{-0.7cm}
\Statex \comm{Compute the final conformance cost}
\State $\ell_{\mathrm{C}}^* \gets \gamma \alpha^*$
\end{algorithmic}
\end{algorithm}

\begin{theorem}[White-Box Identification of Uncertainty Sets]\label{theo:White}
\cref{prob:idWhite} can be solved with \cref{alg:white}.
    
    \begin{proof}             
    If the sets of linearization errors $\widehat{\mathcal{E}}_k^{(m)}$ used in \cref{eq:reachability} contain the zero vectors, i.e., $\mathbf{0}\in \widehat{\mathcal{E}}_k^{(m)}$, $k=0,\dots,n_{\mathrm{k}}-1$, then $\widehat{\Reach}_k^{(m)}$ is underapproximated by $\bar{\Reach}_k^{(m)}$, which can be computed using \cref{eq:reachability_GLO}.    
    By using this underapproximation and defining the function
    $r(\widehat{\Reach}_{k}^{(m)}) = \|  \bar{\Reach}_k^{(m)} \|_I$ as in \cref{lem:cost},
    \cref{prob:idWhite} can be simplified to
    \begin{subequations}
    \begin{align}
        &\argmin_{\alpha, c_{\mathrm{\Delta}}} \quad \gamma \alpha \label{eq:conf_cost_rel}\\
        \text{s.t. } &\quad \forall m, k,  s\colon y^{(m,s)}_k(\sysT) \in \bar{\Reach}_k^{(m)}. \label{eq:conf_constr_rel}
    \end{align}
    \end{subequations} 
    Feasible solutions to this optimization problem lead to reachset-conformant models since the satisfaction of the more restrictive constraint \cref{eq:conf_constr_rel} implies the satisfaction of \cref{eq:conf_constr}.  
    According to \cref{lem:half} and \cref{lem:gen}, \cref{eq:conf_constr_rel} can be expressed with the linear constraints given in \cref{eq:contConstraint} and \cref{eq:posconstraint}.

    The requirement $\mathbf{0}\in \widehat{\mathcal{E}}_k^{(m)}$ is trivially fulfilled for linear systems. For nonlinear systems, this condition is satisfied by \cref{cor:conv_ss} and \cref{cor:conv_io} 
    if the initial state and inputs of the reference trajectory $(\bar{x}_{0}^{(m)},\bar{u}_{0:k}^{(m)})$, which are used for the linearization, are contained in the initial state and input sets, respectively, i.e., $\bar{x}_{0}^{(m)} \in x_{*,0}^{(m)}\oplus \uncX^{(m)}$ and $\bar{u}_{i}^{(m)} \in u_{*,i}^{(m)}\oplus \uncU{i}$, $\forall i$.
    Since $x_{*,0}^{(m)}\oplus \uncX^{(m)}$ and $u_{*,i}^{(m)}\oplus \uncU{i}^{(m)}$ have their centers at $\bar{x}_{0}^{(m)}+ c_{\mathrm{\Delta},\mathrm{x}}$ and $\bar{u}_i^{(m)}+c_{\mathrm{\Delta},\mathrm{u}}$, this containment can be satisfied by setting $c_{\mathrm{\Delta}} = [c_{\mathrm{\Delta},\mathrm{x}}^\top~c_{\mathrm{\Delta},\mathrm{u}}^\top]^\top$ to the zero vector with \cref{eq:deltaCconstraint}.
	\end{proof}
\end{theorem}



\section{Identification of Gray-Box Models}
\label{sec:idGray}
In this section, we extend the uncertainty identification approach from \cref{theo:White} to the reachset-conformant identification of gray-box models (\cref{prob:idGray}).
First, we describe an algorithm that integrates the LP from~\cref{sec:idWhite} in a nonlinear program (NLP) to simultaneously identify the uncertainty sets and the unknown model parameters.
As an alternative with lower computational complexity, we present a sequential identification approach, which identifies the model parameters prior to the uncertainty estimation. 


The uncertainty sets and unknown model parameters $p$ can be simultaneously identified with the approach proposed in~\cite[Sec.~III-B]{liu2021conf}, termed \texttt{GraySim}.
The general procedure is provided in~\cref{alg:idSim}. 
The function \texttt{White} implements \cref{alg:white} and returns scaling factors $\alpha$, center shifts $c_{\mathrm{\Delta}}$, and the conformance cost $\ell_{\mathrm{C}}$ for the {model functions $\func_p$}.
The outer nonlinear program uses $\ell_{\mathrm{C}}$ to optimize the parameters $p$ in \cref{line:graySim1,line:graySim2,line:graySim3,line:graySim4} until the solver converges or exceeds a maximum number of iterations. 
While we do not have any guarantees for finding the globally optimal parameters, we can ensure that the final model is reachset-conformant by computing the scaling factors $\alpha^*$ and center shifts $c_{\mathrm{\Delta}}^*$ with \texttt{White} for the final parameters $p^*$ in \cref{line:graySim_whiteFinal}.

\begin{algorithm}[bt]
\caption{\texttt{GraySim}.}\label{alg:idSim} 
\begin{algorithmic}[1]
\Statex \textbf{Inputs:}
  \Desc{~~$p$}{Initial estimate for the model parameters} 
  {\Desc{~~$\func_p$}{Model functions parameterized by p} 
  \Desc{~~$B_c$}{Center identification matrix}} 
  \Desc{~~$\mathcal{M}$}{Test cases}
\Statex \textbf{Outputs:}
  \Desc{~~$p^*$}{Optimized model parameters}
  \Desc{~~$\alpha^*$}{Optimized scaling factors}
  \Desc{~~$c_{\mathrm{\Delta}}^*$}{Optimized center shifts}
  \Desc{~~$\ell_{\mathrm{C}}^*$}{Final conformance cost}  
\vspace{0.2cm}
\Statex \comm{Find parameter $p$ that minimize the cost of the white-box identification}
\While{NLP solver has not terminated} \label{line:graySim1}
\State $\alpha$, $c_{\mathrm{\Delta}},\ell_{\mathrm{C}} \gets$~\texttt{White}(${\func_p,B_c},\mathcal{M}$)\label{line:graySim2}
\State $p \gets$ next NLP-iteration to minimize $\ell_{\mathrm{C}}$\label{line:graySim3}
\EndWhile\label{line:graySim4}
\State $p^* \gets p$
\Statex \comm{Optimize scaling factors, center shifts, and cost given $p^*$}
\State $\alpha^*,c_{\mathrm{\Delta}}^*,\ell_{\mathrm{C}}^* \gets$~\texttt{White}(${\func_{p^*},B_c},\mathcal{M}$) \label{line:graySim_whiteFinal}
\end{algorithmic}
\end{algorithm}


Solving the LP in \texttt{White} at each iteration of the NLP, as proposed in~\cite{liu2021conf}, leads to high computational complexity. 
Furthermore, only NLP solvers that do not require a closed-form expression of the cost function can be used. 
Thus, we propose \cref{alg:idSeq}, termed \texttt{GraySeq}, resulting in shorter computation times and higher flexibility in the choice of the NLP solver. 
First, we identify model parameters $p^*$ and center shifts $c_{\mathrm{\Delta}}$ in \cref{line:graySeq1,line:graySeq2,line:graySeq7,line:graySeq8,line:graySeq9,line:graySeq10} by minimizing an approximation of the conformance cost $\ell_{\mathrm{C}}$, which is independent of the scaling factors $\alpha$.
Next, we {run} the white-box identification algorithm \texttt{White} {using the optimized parameters $p^*$} to compute the uncertainty sets and the conformance cost in \cref{line:graySeq_whiteFinal}.

\begin{algorithm}[bt]
\caption{\texttt{GraySeq}.}\label{alg:idSeq} 
\begin{algorithmic}[1]
\Statex \textbf{Inputs:}
  \Desc{~~$p$}{Initial estimate for the model parameters}
  {\Desc{~~$\func_p$}{Model functions parameterized by p} 
  \Desc{~~$B_c$}{Center identification matrix}} 
  \Desc{~~$\mathcal{M}$}{Test cases}
\Statex \textbf{Outputs:}
  \Desc{~~$p^*$}{Optimized model parameters}
  \Desc{~~$\alpha^*$}{Optimized scaling factors}
  \Desc{~~$c_{\mathrm{\Delta}}^*$}{Optimized center shifts}
  \Desc{~~$\ell_{\mathrm{C}}^*$}{Final conformance cost}  
\vspace{0.2cm}
\Statex \comm{Find parameter $p$ that minimize the cost function from \cref{prop:costUnder}}
\While{NLP solver has not terminated}\label{line:graySeq1}
    \Statex \quad\, \comm{Compute GLO approximation for each test case}   
    \State {$\bar{\tau}^{(m)} \gets$ \texttt{computeGLO}$(\func_p, m), \forall m\in\mathcal{M}$}\label{line:graySeq2}
\State $\ell_{\mathrm{U}} \gets $ \cref{eq:costUnder} \label{line:graySeq7}
\State $p, c_{\mathrm{\Delta}} \gets$ next NLP-iteration to minimize $\ell_{\mathrm{U}}$\label{line:graySeq8}
\EndWhile\label{line:graySeq9}
\State $p^* \gets p$\label{line:graySeq10}
\Statex \comm{Optimize scaling factors, center shifts, and cost given $p^*$}
\State $\alpha^*,c_{\mathrm{\Delta}}^*,\ell_{\mathrm{C}}^* \gets$~\texttt{White}(${\func_{p^*},B_c},\mathcal{M}$)\label{line:graySeq_whiteFinal}
\end{algorithmic}
\end{algorithm}

We propose to use the following cost function for the estimation of the parameters $p$:

\begin{proposition}[Underapproximative Cost Function] \label{prop:costUnder}
    The minimum cost $\gamma \alpha^*$ of the LP in \cref{eq:conf_opt} can be underapproximated by
    \begin{align} 
        \ell_{\mathrm{U}} &= \sum_{m\in\mathcal{M}} \sum_{k=0}^{n_{\mathrm{k}}-1} w_k \mathbf{1}^\top \mathrm{max}_s
         |y_{\mathrm{a},k}^{(m,s)}-{y}_{\mathrm{\Delta},k}^{(m)}|, \label{eq:costUnder}
    \end{align}
    with 
    {\begin{align*}
        y_{\mathrm{a},k}^{(m,s)}&= y_{k}^{(m,s)}(\sysT)-  \bar{y}_k^{(m)}, \\ {y}_{\mathrm{\Delta},k}^{(m)}&=\begin{bmatrix}
        	\bar{C}^{(m)}_k & \sum_{i=0}^k \bar{D}_{k,i}^{(m)}
            \end{bmatrix}\tilde{B}_c c_{\mathrm{\Delta}},\\
        \tilde{B}_c &=\begin{cases}
        \mathbf{0} & \text{if any function in $\func_p$ is nonlinear},\\
            |B_c-\mathbf{I}|  & \text{else}.
       \end{cases}
    \end{align*}}
    
    \begin{proof}
        Given \cref{lem:gen}, we have $|\beta^{(m,s)}_{k}| \leq \alpha_k^*$ (see \cref{eq:beta1}) with $\alpha_k^*\coloneqq[\alpha_{\mathrm{x}}^{*\top} ~\alpha_{\mathrm{u}}^{*\top}  ~\cdots~\alpha_{\mathrm{u}}^{*\top} ]^\top$ and
        \begin{align*} 
             y_{\mathrm{a},k}^{(m,s)} - & \bar{C}^{(m)}_k c_{\mathrm{\Delta},\mathrm{x}} -\sum_{i=0}^k \bar{D}_{k,i}^{(m)}c_{\mathrm{\Delta},\mathrm{u}} = \mathrm{gen}'\bigl({\bar{\Reach}_{k}^{(m)}}\bigr) \beta^{(m,s)}_{k}.
        \end{align*}
        By taking the absolute value of both sides, we obtain
        \begin{align*} 
              |y_{\mathrm{a},k}^{(m,s)} -  \bar{C}^{(m)}_k c_{\mathrm{\Delta},\mathrm{x}} -\sum_{i=0}^k \bar{D}_{k,i}^{(m)} &c_{\mathrm{\Delta},\mathrm{u}}| 
              = |\mathrm{gen}'\bigl({\bar{\Reach}_{k}^{(m)}}\bigr) \beta^{(m,s)}_{k}| \\
             &\leq |\mathrm{gen}'\bigl({\bar{\Reach}_{k}^{(m)}}\bigr)| ~|\beta^{(m,s)}_{k}| \\
             &\leq |\mathrm{gen}'\bigl({\bar{\Reach}_{k}^{(m)}}\bigr)| ~\alpha_k^*. 
        \end{align*}        
        As this has to hold for all $s$, we can take the maximum over $s$, resulting in
        \begin{align} 
              \max_s |y_{\mathrm{a},k}^{(m,s)} - & \bar{C}^{(m)}_k c_{\mathrm{\Delta},\mathrm{x}} -\sum_{i=0}^k \bar{D}_{k,i}^{(m)}c_{\mathrm{\Delta},\mathrm{u}}|\leq |\mathrm{gen}'\bigl({\bar{\Reach}_{k}^{(m)}}\bigr)| ~\alpha_k^*. \notag 
        \end{align}        
        At the same time, the minimum cost of \cref{eq:conf_opt} can be computed as
        \begin{align*}
            \ell_{\mathrm{C}}^* 
            &= \sum_{m\in\mathcal{M}} \sum_{k=0}^{n_{\mathrm{k}}-1} w_k \mathbf{1}^\top |\mathrm{gen}'(\bar{\Reach}_{k}^{(m)})|~\alpha_k^*.
        \end{align*}
        Combining the last two formulas and considering constraint \cref{eq:deltaCconstraint} leads to the underapproximation $ \ell_{\mathrm{U}}\leq \gamma \alpha^*$.        
    \end{proof}
\end{proposition}

Although we do not have any guarantees on the closeness of the conformance cost $\ell_{\mathrm{C}}$ and the underapproximation $\ell_{\mathrm{U}}$, this cost function works well in practice as demonstrated in \cref{sec:experiments}.
Alternatively, we can use the least-squares cost function 
\begin{align}
	\ell_{\mathrm{LS}} = \sum_{m\in\mathcal{M}} \sum_{k=0}^{n_{\mathrm{k}}-1} w_k \mathbf{1}^\top \sum_{s=1}^{n_{\mathrm{s}}}
	\left(y_{\mathrm{a},k}^{(m,s)}-{y}_{\mathrm{\Delta},k}^{(m)}\right)^2. \label{eq:lsCost}
\end{align} 
This cost function decreases the computation time even more, but the results are more vulnerable to not centrally distributed uncertainties compared to using cost functions that consider only the maximum errors.
A detailed analysis of the accuracy and computation times is presented in \cref{sec:exp_gen}.


\section{Identification of Black-Box Models}
\label{sec:idBlack}

For the black-box identification of reachset-conformant models (\cref{prob:idBlack}), we propose a sequential approach that first identifies {the model functions followed by the computation of the uncertainty sets using the methods from \cref{sec:idWhite}.}
We use genetic programming (GP) in this work, but other black-box identification methods, such as the training of neural networks, can be used similarly \cite{mohajerin2019rnnSI,nelles1996basisNN,suykens1995neuralSS,sjoeberg1995blackbox}.

\begin{figure}[t]
    \centering
    \hspace{-0.3cm}
    \includegraphics[width=0.5\linewidth, keepaspectratio,page=10]{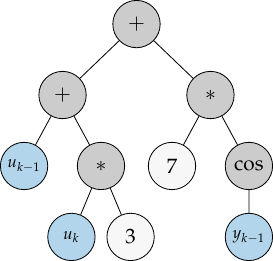}
    \caption{Example gene in GP using the building blocks $+$, $*$, and $\cos$, the input variables $u_{k}$, $u_{k-1}$, and $y_{k-1}$, and the random constants $3$ and $7$.}
    \label{fig:gp}
\end{figure}

GP is a global optimization technique that iteratively evolves a population of symbolic functions or entire computer programs \cite{koza1994gp}.
The evolution is driven by the need to optimize a fitness function mimicking the Darwinian principle of survival of the fittest. 
GP for system identification can be described by the following steps:
\begin{enumerate}
    \item Initialization: A population of random symbolic functions, called individuals, is generated as a weighted sum of different genes. Each gene is a nonlinear function represented by a tree structure, constructed using a specified set of building block functions, input variables, and optionally, random constants. An example gene is displayed in \cref{fig:gp}.
    \item Fitness Evaluation: The fitness function evaluates the fit of each individual to the training data. 
    \item Evolution: Individuals are selected based on their fitness alone or on both their fitness and their complexity. The selected individuals undergo the genetic operations crossover (genes are exchanged between individuals), mutation (random changes to individuals to maintain genetic diversity), and replication (some high-performing individuals are copied directly to the next generation).
    The resulting individuals are part of the next generation, where we continue with step 2.
    \item Termination: The evolution process is terminated if a fitness threshold is exceeded or the maximum number of generations is reached.
\end{enumerate}
Overfitting to the given data is mitigated by limiting the maximum number of nodes and the maximum depth of the trees, as well as penalizing complexity in the evolution process. 

\begin{algorithm}[h!]
\caption{\texttt{BLackCGP}.}\label{alg:blackCGP} 
\begin{algorithmic}[1]
		\Statex \textbf{Inputs:}
		\Desc{~~$\mathcal{M}_{\mathrm{T1}}$}{Training test cases for step 1}
		\Desc{~~$\mathcal{M}_{\mathrm{T2}}$}{Training test cases for step 2}
		\Desc{~~$\mathcal{M}_{\mathrm{C}}$}{Conformance test cases}
		\Statex \textbf{Outputs:}
		{\Desc{~~$\func^*$}{Identified model functions}}
		\Desc{~~$\alpha^*$}{Optimized scaling factors}
		\Desc{~~$c_{\mathrm{\Delta}}^*$}{Optimized center shifts}
            \Desc{~~$\ell_{\mathrm{C}}^*$}{Final conformance cost}  
		\vspace{0.2cm}
    \Algphase{Step 1: Minimize least-squares cost} \vspace{-0.1cm}
    \Statex \comm{Initialize the function sets}
    \State ${\{\func_{1},\dots, \func_{n_{\mathrm{f}}} \}} \gets$      \texttt{Initialize} \label{line:init}
    \For{$i=\{1,\dots, n_{\mathrm{g}}\}$} \label{line:start_gs}
        \Statex \quad\, \comm{Compute least-squares cost for each function set}
	\For{$j=\{1,\dots, n_{\mathrm{f}}\}$}
            \State {\small $\bar{\tau}^{(m)} \gets$ {\texttt{computeGLO}$(\func_j, m), \forall m\in\mathcal{M}_{\mathrm{T1}}$}\label{line:CDcomp2}
		\State  $\ell_j \gets $ \cref{eq:lsCost}}
	\EndFor
        \If{$i<n_{\mathrm{g}}$}
            \Statex \quad\quad~~~\comm{Optimization via genetic evolution}
		\State {{\small$\{\func_{1},\dots, \func_{n_{\mathrm{f}}} \} \gets$ \texttt{Evolve}$\bigl((\func_{1},\ell_1),\dots,(\func_{n_{\mathrm{f}}},\ell_{n_{\mathrm{f}}})\bigr)$}}
        \EndIf
    \EndFor \label{line:end_gs}
    \Statex \comm{Selection of the $\tilde{n}_{\mathrm{f}}$ function sets with the lowest cost}
	\State {$\{\func_{1},\dots, \func_{\tilde{n}_{\mathrm{f}}} \} \gets$ \texttt{Best}$\bigl((\func_{1},\ell_1),\dots,(\func_{n_{\mathrm{f}}},\ell_{n_{\mathrm{f}}})$, $\tilde{n}_{\mathrm{f}}\bigr)$} \label{line:best_gs}
  
    \Algphase{Step 2: Minimize conformance cost}
    \For{$i=\{1,\dots, \tilde{n}_{\mathrm{g}}\}$}		
        \Statex \quad\, \comm{Compute sum of conformance costs for each function set}
	\For{$j=\{1,\dots, \tilde{n}_{\mathrm{f}}\}$}\label{line:start_DV}
            \State {\small$\ell_{j} \gets 0$ \label{line:start_DV00}}
		\For{$l=\{1,\dots,n_v\}$} \label{line:start_DV0} 
                \State {\small$\alpha$, $c_{\mathrm{\Delta}},\ell_{\mathrm{C}} \gets$~\texttt{White}(${\func_j},\mathbf{0},\mathcal{M}_{\mathrm{T2},l}$)\label{line:start_DV1} 
		      \State $\ell_{j}  \gets \ell_{j} + \ell_{\mathrm{C}}$} \label{line:start_DV2} 
		\EndFor \label{line:start_DV3}
	\EndFor\label{line:start_DV4}
        \If{$i<\tilde{n}_{\mathrm{g}}$}
            \Statex \quad\quad~~~\comm{Optimization via genetic evolution}
            \State {{\small$\{\func_{1},\dots, \func_{\tilde{n}_{\mathrm{f}}} \} \gets$ \texttt{Evolve}$\bigl((\func_{1},\ell_1),\dots,(\func_{\tilde{n}_{\mathrm{f}}},\ell_{\tilde{n}_{\mathrm{f}}})\bigr)$}\hspace{-0.2cm}}
        \EndIf
    \EndFor
    \Statex \comm{Selection of the function set with the lowest cost}
    \State {$\func^* \gets$ \texttt{Best}$\bigl((\func_{1},\ell_1),\dots,(\func_{\tilde{n}_{\mathrm{f}}},\ell_{\tilde{n}_{\mathrm{f}}
		})$, $1\bigr)$} \label{line:best}
    \Statex \comm{Optimize scaling factors, center shifts, and cost given $\func^*$}
    \State $\alpha^*,c_{\mathrm{\Delta}}^*,\ell_{\mathrm{C}}^* \gets$~\texttt{White}(${\func^*},\mathbf{0},\mathcal{M}_{\mathrm{C}}$) \label{line:opt} 
\end{algorithmic}
\end{algorithm}

As some of the models identified with GP might lead to exploding reachable sets, even though we observed small prediction errors in the training process, we propose a new GP variant termed \emph{conformant genetic programming (CGP)}. 
Exploding reachable sets can come from inherent model dynamics that were not observable for the test cases in the training dataset.
For example, we might have a division by a near-zero number in the output equation for a particular input trajectory, which would lead to large outputs.
The probability that this input trajectory is contained in the training data is small, but the probability is significantly higher when considering whole input sets. 
Thus, it is advantageous to consider the cost $\ell_{\mathrm{C}}$ of \cref{eq:conf_opt}, which describes the size of the reachable set for a conformant model, in the training process.
However, since the computation of the conformance cost requires the solution of an LP, this cost leads to a significantly higher computational complexity than a least-squares cost function like \cref{eq:lsCost} and can only be evaluated for small test suites in reasonable computation time.
Hence, we propose the approach described in \cref{alg:blackCGP}, where we first minimize a least-squares cost function, followed by a few iterations of minimizing the conformance cost.
We use the following functions in the algorithm:
\begin{itemize}
        \item \texttt{Initialize}, which randomly creates $n_{\tau}$ {sets of model functions} with different nonlinearities \cite[Sec.~3.2]{searson2015gptips},
	\item \texttt{Evolve}, which implements the evolution process based on the costs of the {function sets} \cite[Sec.~3.2]{searson2015gptips},
	\item \texttt{Best}, which returns the {sets of model functions} with the lowest cost, where the last argument specifies the number of returned {sets}, and
	\item \texttt{White}, which calls \cref{alg:white}.
\end{itemize}

First, we randomly initialize a population of $n_{\mathrm{f}}$ {sets of model functions, where the structure of each set $\func_j$ depends on the selected model class (e.g., $\func_j=\{h_j\}$ from \cref{eq:narx} for input-output models)}.
For the first $n_{\mathrm{g}}$ generations, we minimize the least-squares cost function for the training dataset $\mathcal{M}_{\mathrm{T1}}$ to efficiently find good candidate functions (see \cref{line:start_gs} to \cref{line:end_gs}).
From the optimized {function sets}, we choose the $\tilde{n}_{\mathrm{f}}$ {sets} with the lowest costs in \cref{line:best_gs}.
In the second part of the algorithm, we optimize the obtained functions by minimizing the conformance cost for the second training dataset $\mathcal{M}_{\mathrm{T2}}$.
To keep the computational complexity small, we split the dataset $\mathcal{M}_{\mathrm{T2}}$ into the subdatasets $\mathcal{M}_{\mathrm{T2},l}$, $l=1,\dots,n_v$, and compute the sum of conformance costs over all subdatasets  for each function set in \cref{line:start_DV,line:start_DV00,line:start_DV0,line:start_DV1,line:start_DV2,line:start_DV3,line:start_DV4}. 
{To activate the identification of center shifts in the rare case of linear model functions $\func_j$, we initialize $B_c$ with the zero matrix (the algorithm \texttt{White} will automatically set $B_c$ to the identity matrix for nonlinear models).}
Having reached the maximum number of generations $\tilde{n}_{\mathrm{g}}$, we choose the {function set} with the lowest cost in \cref{line:best} and compute the scaling factors and center shifts of a reachset-conformant model for the test cases in $\mathcal{M}_{\mathrm{C}}$ in \cref{line:opt}.


\section{Numerical Experiments}
\label{sec:experiments}
First, we evaluate the scalability of our white-box identification approach.
Next, we analyze the general performance of the proposed white-, gray-, and black-box identification methods for different model types and compare them to baseline methods. 
Finally, we identify a reachset-conformant vehicle model from real-world measurements.

All methods are implemented in the MATLAB toolbox CORA \cite{althoff2015introduction} and are evaluated on an AMD EPYC 7763 processor with 2TB RAM and NVIDIA A100 40GB GPU.
We use the built-in solvers linprog and fmincon for solving the required LPs and NLPs.
Continuous-time models will be discretized for the identification using the forward Euler method.

\subsection{Scalability of the White-Box Identification Approach} \label{sec:exp_scal}

For evaluating the scalability of \cref{alg:white}, we simulate a discretized version of a water tank system with uncertain inputs~\cite{althoff2008nonlin}, whose dynamics are
\begin{align*}
    \dot{x}_{t,1} &= u_{t,1} - 0.3\sqrt{x_{t,1}}\\
    \dot{x}_{t,i} &= u_{t,i} + 0.3(\sqrt{x_{t,i-1}} - \sqrt{x_{t,i}}),\quad i=2,\dots,n_{\mathrm{x}}.
\end{align*}
The state $x_{t,i}$ describes the water level of tank $i$ at time $t$, such that the system dimension $n_{\mathrm{x}}$ is equal to the number of tanks, and $u_{t,i}$ the water flow into tank $i$.
We assume that there is an inflow in only every third tank, $u_{t,i}=0$ for $i\not=1,4,7,\dots$, and we measure each state variable, i.e., $y_{t,i} = x_{t,i}$ for $i=1,\dots,n_{\mathrm{y}}$.
The computation times of the LP from \cref{theo:White} using the halfspace constraints from~\cref{lem:half} and the generators constraints from~\cref{lem:gen} are presented for a) varying time horizons $n_{\mathrm{k}}$, b) a varying number of test cases, which is denoted by $n_{\mathrm{m}}$, c) a varying number of sample executions $n_{\mathrm{s}}$ of each test case, and d) a varying measurement dimension $n_{\mathrm{y}}$.
As we can see from~\cref{tab:compTimes}, increasing the number of samples $n_{\mathrm{s}}$ does not influence the computational complexity of the LP using the halfspace constraints since we only consider the sample $y_{\mathrm{a}}^{(m,s)}$ closest to each halfspace in \cref{eq:conf_constr_half1}.
However, as the halfspace conversion scales exponentially with the dimension of the reachable set \cite{luetzow2024generator}, halfspace constraints can only be used for systems with small measurement dimension $n_{\mathrm{y}}$ and small time horizons $n_{\mathrm{k}}$. In contrast, using the generator constraints is efficient for high-dimensional systems and long time horizons, but its computational complexity depends substantially on the overall number of measurement trajectories $n_{\mathrm{s}}\cdot n_{\mathrm{m}}$.

\begin{table}[bt]
    \caption{Computation times for white-box identification.}
    \label{tab:compTimes}
    \centering
    \begin{tabular}{c c c c c c }
       \toprule
       \multicolumn{4}{c}{} & \multicolumn{2}{c}{\textbf{Computation time of \cref{theo:White} [s]}} \\
       $n_{\mathrm{k}}$ & $n_{\mathrm{m}}$ & $n_{\mathrm{s}}$ & $n_{\mathrm{y}}$ & {Using \cref{lem:half}}& {Using \cref{lem:gen}} \\
       \cmidrule(lr){1-4}\cmidrule(lr){5-6}
        4 & 10 & 1 & 3 & 0.16 & \textbf{0.15}  \\
        20 & 10 & 1 & 3 & 16.14 & \textbf{0.83} \\
        40 & 10 & 1 & 3 & $>$600.00 & \textbf{1.61} \\
        4 & 100 & 1 & 3 & 0.54 & \textbf{0.26}  \\ 
        4 & 1000 & 1 & 3 & 23.26 & \textbf{2.58}  \\
        4 & 10 & 10 & 3 & \textbf{0.16} & 0.18  \\
        4 & 10 & 100 & 3 & \textbf{0.16} & 1.43  \\
        4 & 10 & 1 & 6 & 9.84 & \textbf{0.22}  \\
        4 & 10 & 1 & 9 & $>$600.00 & \textbf{0.31}  \\        
        4 & 10 & 1 & 100 & $>$600.00 & \textbf{8.53}  \\
        \bottomrule
    \end{tabular}    
\end{table}

\subsection{General Comparison for Different Model Types}  \label{sec:exp_gen}
We evaluate the proposed identification methods by computing reachset-conformant models for four dynamical systems, which are 
\begin{itemize}
    \item a pedestrian model \cite{luetzow2023armax} formulated as the {linear state-space model} 
    \begin{subequations}\label{eq:pedSS}
     \begin{align*}
     x_{k+1} &= \begin{bmatrix}p_1 & 0 & p_2 & 0 \\
    0 & p_1 & 0 & p_2\\
    0 & 0 & p_1 & 0 \\
    0 & 0 & 0 & p_1 
    \end{bmatrix} x_k + \begin{bmatrix}p_3 & 0 & 0 & 0\\
    0 & p_3 & 0 & 0\\ 
    p_4 & 0 & 0 & 0\\ 
    0 & p_4 & 0 & 0
    \end{bmatrix} u_k,\\
    y_k &= \begin{bmatrix} 1 & 0 & 0 & 0 \\
    0 & 1 & 0 & 0
    \end{bmatrix}x_k + \begin{bmatrix} 0 &  0 & 1 & 0\\
    0 &  0 & 0 & 1\end{bmatrix} u_k,
    \end{align*}        
    \end{subequations}
    where $p_1=1$, $p_2=0.01$, $p_3=5\cdot 10^{-5}$, $p_4 = 0.01$, 
    \item the pedestrian model converted to an {ARX model}, using the conversion formulas from \cite[Prop.~2]{luetzow2023armax}, where
    \begin{align*}
    y_k &= \begin{bmatrix}
            p_1 & 0 \\
            0 & p_1
        \end{bmatrix} y_{k-1} + \begin{bmatrix}
            p_2 & ~0 \\
            ~0 & p_2
        \end{bmatrix} y_{k-2} \\
         &\quad+ \begin{bmatrix}
            ~0 & ~0 & ~1 & ~0 \\
            ~0 & ~0 & ~0 & ~1
        \end{bmatrix} u_k  + \begin{bmatrix}
            p_3 & 0 & p_4 & 0 \\
            0 & p_3 & 0 & p_4
        \end{bmatrix} u_{k-1} \\ 
        &\quad + \begin{bmatrix}
            p_3 & 0 & ~1 & ~0 \\
            0 & p_3 & ~0 & ~1
        \end{bmatrix}u_{k-2},
    \end{align*}
    with $p_1=2$, $p_2=-2$, $p_3=5\cdot 10^{-5}$, $p_4 = -1$,
    \item the Lorenz system~\cite{lorenz1963} described by the {nonlinear state-space model} 
    \begin{subequations}
    \begin{align*}
    \dot{x}_t &= \begin{bmatrix}
        \bigl(p_1+u_{t,1}\bigr) \bigl(x_{t,2}-x_{t,1}\bigr) \\
        \bigl(p_2+u_{t,2}\bigr) x_{t,1}-x_{t,2}-x_{t,1}x_{t,3} \\
        x_{t,1}x_{t,2}-\bigl(p_3+u_{t,3}\bigr)x_{t,3}
    \end{bmatrix}\\
    y_t &= \begin{bmatrix}
        x_{t,1}\\
        x_{t,2}
    \end{bmatrix},
    \end{align*}
    \end{subequations}
    with $p_1=10$, $p_2=28$, $p_3=\frac{8}{3}$,
    \item a {NARX model}, which we call NARX1 and whose dynamics are adapted from \cite{kroll2014benchmark}
    \begin{align*}
        y_k = \begin{bmatrix}
            \dfrac{y_{k-1,1}}{1+y_{k-1,2}y_{k-1,2}} + p_1 u_{k-1,1}\\
            \dfrac{\vphantom{W^X} y_{k-1,1}y_{k-1,2}}{1+y_{k-1,2}y_{k-1,2}} + p_2 u_{k-2,2}
        \end{bmatrix},
    \end{align*}
    with $p_1=0.8$, $p_2=1.2$.
\end{itemize}
The corresponding uncertainty sets $\uncX^{(m)}$ (only for the state-space models) and $\uncU{i}^{(m)}$ are represented by zonotopes with a random center vector with elements between $-1$ and $1$ and a random diagonal generator matrix with elements between $-0.25$ and $0.25$.

Using the true model, the true uncertainty sets, and random input trajectories, we simulate 100 conformance test suites $\mathcal{M}_{\mathrm{C}}$ comprising $n_{\mathrm{m}}=20$ test cases each. Each test case has a length of $n_{\mathrm{k}}=n_{\mathrm{p}}+6$ time steps and contains $n_{\mathrm{s}}=10$ measurement trajectories resulting from different uncertainty trajectories, where $n_{\mathrm{p}}$ is the model order for the ARX and NARX model (we require $n_{\mathrm{p}}$ measurements for their initialization) and zero for the state-space models if not defined otherwise.
More data is required for the black-box identification approach as explained in the respective subsection. 
The matrices $G_{\mathrm{x}}$ and $G_{\mathrm{u}}$ for the uncertainty sets are initialized by identity matrices, and the initial estimates for the unknown parameters and center vectors are sampled from a Gaussian distribution with a standard deviation of 0.01.

We compare the identification methods using the following criteria:
\begin{itemize}
    \item the \emph{normalized cost}, which is the ratio of the final cost $\ell_{\mathrm{C}}^*$ of the model identified with the considered approach and the cost using the true model and uncertainty sets, 
    \item the \emph{failure rate}, which is the ratio of the number of failure test suites, i.e., test suites where a conformant model could not be identified or where the normalized cost is bigger than 100, to the total number of test suites,
    \item the \emph{computation time}, which is the time required for solving the conformance problem for one test suite, and
    \item the root mean squared errors (\emph{RMSE}) between the true parameters and the parameters estimated with the gray-box identification approaches. 
\end{itemize} 
In \cref{tab:comp}, we present the results averaged over all dynamical systems and non-failure test suites (except for the failure rate) for the proposed identification methods and baseline methods.
Given the absence of comparable solutions in the existing literature, we benchmark our methods \emph{White} from \cref{alg:white}, \emph{GraySeq} from \cref{alg:idSeq}, and \emph{BlackCGP} from \cref{alg:blackCGP} against the following baseline approaches:
\begin{itemize}
    \item \emph{WhiteAdd}: This white-box identification approach is based on \cref{alg:white} but considers only additive measurement error sets in the optimization since previous research focused only on additive uncertainty sets \cite{althoff2012reachability,Liu2023conf}.
    \item \emph{GraySim}: This gray-box identification approach implements the simultaneous identification scheme from \cref{alg:idSim}, which was adapted from \cite{liu2021conf}.
    \item \emph{GraySeq2}: This gray-box identification approach is based on the sequential identification scheme of \emph{GraySeq} in \cref{alg:idSeq} but minimizes the least-squares cost function $\ell_{\mathrm{LS}}$ from \cref{eq:lsCost} with the NLP solver instead of $\ell_{\mathrm{U}}$.
    \item \emph{BlackGP}: This black-box identification approach is based on the standard GP algorithm and aims to find a model minimizing the least-squares cost function in \cref{eq:lsCost} before applying the white-box identification approach to find uncertainty sets that make the model reachset-conformant.
\end{itemize}

\begin{figure}[t]
    \subfloat[Pedestrian state-space model.]{  
    \includegraphics[height=2.5cm, keepaspectratio, page=4]{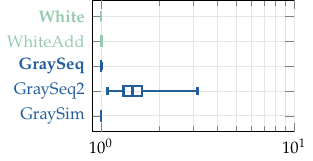}
    \label{fig:nc_lss}
    }
    \subfloat[Pedestrian ARX model.]{  
    \includegraphics[height=2.5cm, keepaspectratio, page=5]{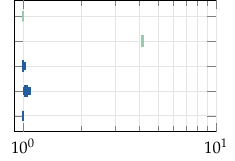}
    \label{fig:nc_arx}
    }
    
    \subfloat[Lorenz model.]{ 
    \includegraphics[height=3.4cm, keepaspectratio, page=6]{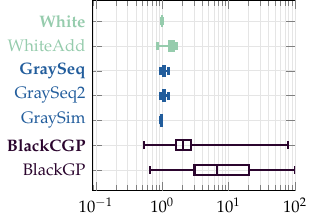}
    \label{fig:nc_nss}
    }
    \subfloat[NARX1 model.]{ 
    \includegraphics[height=3.4cm, keepaspectratio, page=7]{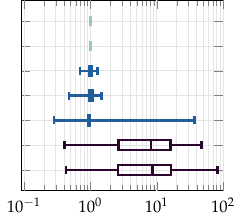}
    \label{fig:nc_narx}
    }
    
    \caption{Normalized cost for the identified models for different non-failure test suites.}
    \label{fig:nc}
\end{figure}

\begin{figure*}[t]
        \centering        
        \includegraphics[width=\textwidth, keepaspectratio, page=3]{figures/2024-02-14_NLIdentification_figure.pdf}
    
    \subfloat[Pedestrian state-space model.]{
        \centering        
        \begin{minipage}{.27\linewidth}
        \includegraphics[width=\linewidth, keepaspectratio]{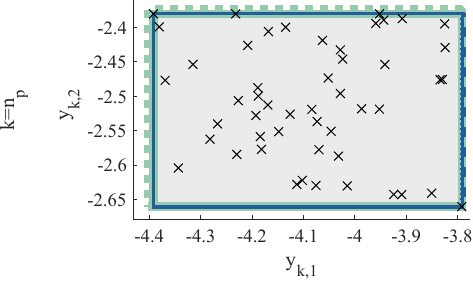}
        \includegraphics[width=\linewidth, keepaspectratio]{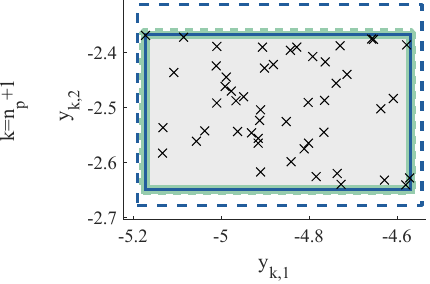}    
        \includegraphics[width=\linewidth, keepaspectratio]{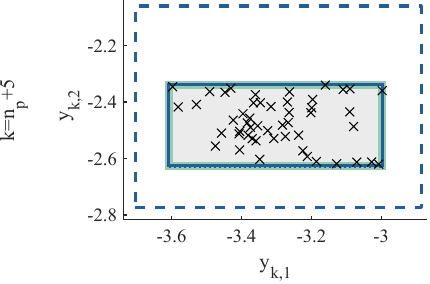}       
        \includegraphics[width=\linewidth, keepaspectratio]{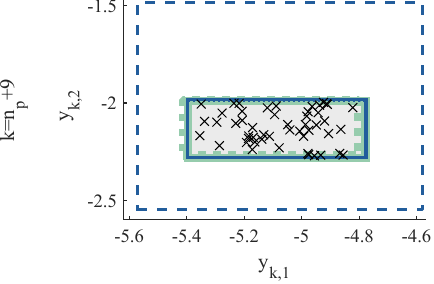}
        \end{minipage}
        \label{fig:lss}
    }
    \subfloat[Pedestrian ARX model.]{
        \centering        
        \begin{minipage}{.22\linewidth}
        \includegraphics[width=\linewidth, keepaspectratio]{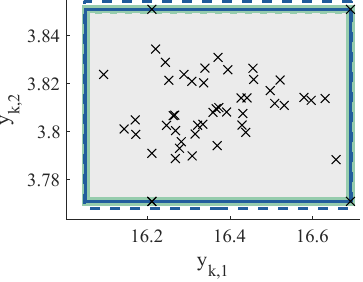}      
        \includegraphics[width=\linewidth, keepaspectratio]{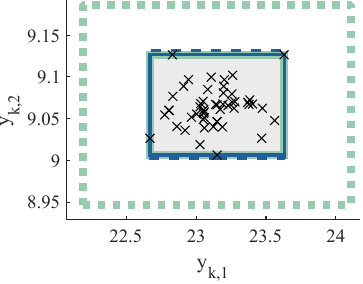}    
        \includegraphics[width=\linewidth, keepaspectratio]{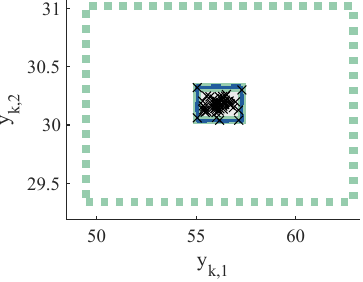}         
        \includegraphics[width=\linewidth, keepaspectratio]{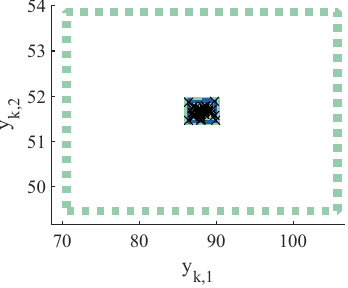}
        \end{minipage}
        \label{fig:arx}
    }
    \subfloat[Lorenz model.]{
        \centering        
        \begin{minipage}{.22\linewidth}
        \includegraphics[width=\linewidth, keepaspectratio]{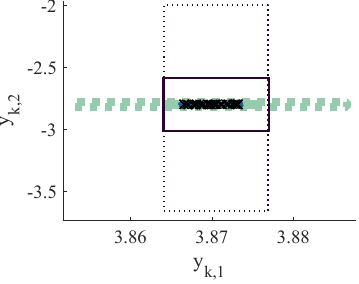}
        \includegraphics[width=\linewidth, keepaspectratio]{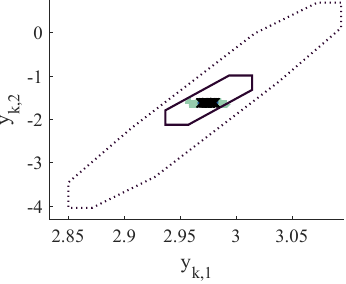}
        \includegraphics[width=\linewidth, keepaspectratio]{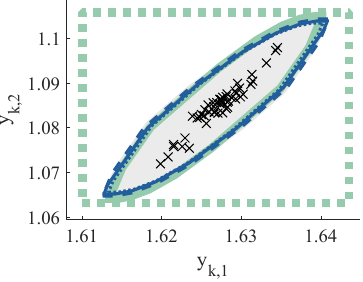}  
        \includegraphics[width=\linewidth, keepaspectratio]{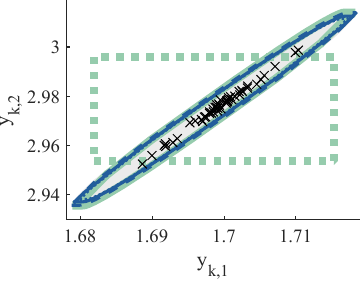}       
        \end{minipage}
        \label{fig:nss}
    }
    \subfloat[NARX1 model.]{
        \centering        
        \begin{minipage}{.22\linewidth}
        \includegraphics[width=\linewidth, keepaspectratio]{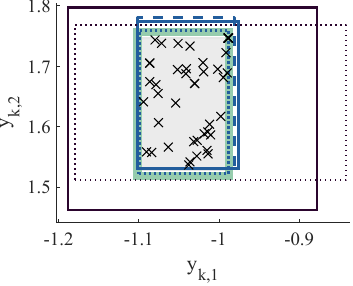}   
        \includegraphics[width=\linewidth, keepaspectratio]{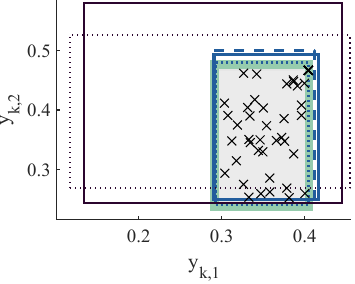}
        \includegraphics[width=\linewidth, keepaspectratio]{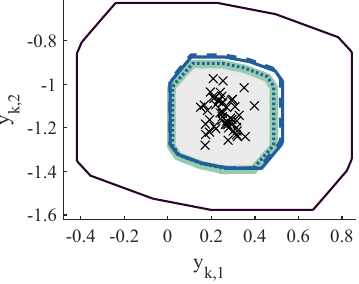}   
        \includegraphics[width=\linewidth, keepaspectratio]{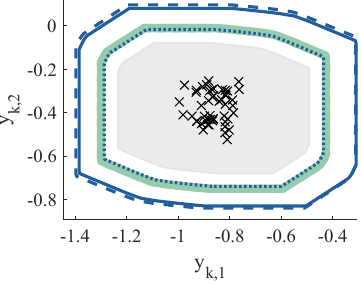}
        \end{minipage}
        \label{fig:narx}
    }
    
    \caption{Example test cases: measurements $y^{(s)}_k$ with $s=1,\dots,50$, reachable set $\mathcal{Y}_{\text{true},k}$ of the true system, and reachable sets $\mathcal{Y}_{\text{method},k}$ from models identified with different white, gray, and black-box identification methods for the time steps $k=n_{\mathrm{p}}, n_{\mathrm{p}}+1, n_{\mathrm{p}}+5, n_{\mathrm{p}}+9$.}
    \label{fig:testcases}
\end{figure*}

As the performance of the identification methods can vary substantially between different test suites and systems, we also visualize the normalized cost as boxplots for all non-failure test suites in \cref{fig:nc}. 
Example test cases, which were not contained in the identification data, consisting of $n_{\mathrm{s}}=50$ output trajectories with a length of $n_{\mathrm{k}}=n_{\mathrm{p}}+10$ time steps, and the reachable sets from the identified models are displayed in \cref{fig:testcases}.
Since the GP approaches approximate the Lorenz model with a NARX model, which requires $n_{\mathrm{p}}=3$ measurements for its initialization, we enforce that all test cases visualized in \cref{fig:nss} start with the same measurements $y_i$, $i=0,\dots,n_{\mathrm{p}}-1$, such that we are able to compute one reachable set, which has to enclose all measurements.
Thus, we start the reachability predictions at the time step $k=n_{\mathrm{p}}$ and, to have a fair comparison, we use $x_{n_{\mathrm{p}}} \oplus \uncX^{(m)}$ as the initial state set,  which contains all states of the visualized test cases at $k=n_{\mathrm{p}}$, for the nonlinear state-space models identified with the white and gray-box approaches.

\begin{table}[t]
	\caption{Evaluation of the proposed identification methods (in \textbf{bold}) and baseline methods.}    
    \label{tab:comp}
    \centering
    \begin{tabular}{c c c c c c}
       \toprule
       \textbf{Approach}   & \textbf{Normalized} &\textbf{Failure}  & \textbf{Computation}  & \textbf{RMSE} \\        
          & \textbf{cost} &\textbf{rate [\%]}  & \textbf{time [s]}  &  \\ 
       \midrule
        \textbf{White}      & 1.00       & 0           &   0.09      &  -    \\ 
        {WhiteAdd}          & 1.88       & 0           &   0.10      &  -    \\ 
        \midrule
        \textbf{GraySeq}	& 1.02       & 4.00        &   12.56     & 0.19  \\ 
        {GraySeq2}            & 1.16       & 0.25        &   4.95      & 0.39  \\ 
        {GraySim}           & 1.49       & 11.50       &   20.67     & 0.24  \\ 
        \midrule
        \textbf{BlackCGP}   & 6.53       & 0.12        &   219.65       &   -   \\ 
        {BlackGP}           & 12.68      & 29.57       &   180.80        &   -  \\
        \bottomrule
    \end{tabular}    
\end{table}

\subsubsection{White-Box Identification}\label{subsec:expWhite}

We use \cref{alg:white} to identify the white-box models. 
Please keep in mind that the ARX and NARX models do not require to find $\alpha_x$ and that $c_{\mathrm{\Delta}}$ is only identified for linear models.
We use the halfspace constraints from \cref{lem:half} instead of the generator constraints from \cref{lem:gen} to shorten computation times, because the considered systems are low-dimensional and the time horizon is short. 

We can see from \cref{tab:comp} that our identification algorithm, termed \emph{White}, is able to estimate small uncertainty sets, which lead to tight reachable sets, in short computation times. 
Since the assumed directions for the generators were correct, linear programming is able to find a solution that is as good as using the true uncertainty sets, i.e., the normalized cost in \cref{fig:nc} is always approximately equal to 1. 
Furthermore, \cref{fig:testcases} shows that the reachable set of the identified model is identical to the reachable set of the true system for the pedestrian state-space model and the pedestrian ARX model.
Due to the linearization error, which is not considered in the identification but added for the computation of the reachable set, \emph{White} identifies more conservative uncertainty sets for the nonlinear models, resulting in larger reachable sets after longer time horizons (see \cref{fig:narx} at $k=n_{\mathrm{p}}+9$). 
The magnitude of the conservatism depends on the degree of the nonlinearities of the model.

The performance of \emph{WhiteAdd} compared to \emph{White} depends on how well additive uncertainties can approximate the uncertainties of the true systems.
The true uncertainties of the pedestrian state-space model and the NARX1 model have a similar influence as additive uncertainties. 
As a result, the normalized cost of the models identified with \emph{WhiteAdd} is almost identical to the cost of the models identified with \emph{White} for these systems (see \cref{fig:nc_lss,fig:nc_narx}).
However, the true uncertainties of the pedestrian ARX model and Lorenz model have a more complex influence on the system, which cannot be well approximated by additive uncertainties.
This can lead to bigger uncertainty sets and reachable sets (see \cref{fig:arx,fig:nss}) and, thus, higher normalized costs (see \cref{fig:nc_arx,fig:nc_nss}).
In some cases, we also obtain models that are not reachset-conformant for validation test cases (see \cref{fig:nss}, where the measurements are not contained in the reachable sets $\mathcal{Y}_{\text{WhiteAdd}}$ at $k=n_{\mathrm{p}}+9$).

If we know how the uncertainties act on the system, we should use this knowledge in the identification by applying the proposed algorithm \emph{White}. 
The computation times are very short, and the identification results are as good as or better than those of an identification algorithm that only considers additive uncertainty sets.

\subsubsection{Gray-Box Identification} \label{subsec:exp_gray}
We evaluate the gray-box identification approaches from~\cref{sec:idGray} by augmenting the identification variables from the previous subsection with the model parameters $p$.
Furthermore, we assume that the center vector of the uncertainty set of the NARX1 model is unknown, such that $c_{\mathrm{u}}$ is added to the identification variables for this model. 

As we can see from \cref{fig:nc}, the proposed sequential approach, termed \emph{GraySeq}, yields the smallest cost values for the linear models and the NARX1 model of all gray-box identification approaches, while the simultaneous identification approach \emph{GraySim}, results in marginally smaller cost values than \emph{GraySeq} for the Lorenz model.
The baseline method \emph{GraySeq2}, which uses a least-squares cost function, leads to the shortest computation times but also higher normalized costs and bigger reachable sets (see \cref{tab:comp,fig:nc,fig:testcases}). 
As displayed in \cref{tab:comp},  all three identification methods lead to small root mean squared errors between the estimated parameters and the true parameters, where \emph{GraySeq} shows the smallest error on average.
\emph{GraySim} has the highest failure rate.

\subsubsection{Black-Box Identification} 

Lastly, we evaluate the black-box identification approaches.
As the model structure of linear models is completely determined by the linearity and the system dimension, which could be estimated by comparing the gray-box identification results for different dimensions, we only apply our black-box identification approaches to the two nonlinear systems.
{We assume no prior knowledge of the system state and thus identify the model function $\func=\{h\}$ of an input-output model, as described in \cref{eq:narx}.}

First, we generate additional training datasets $\mathcal{M}_{\mathrm{T1}}$ consisting of $1500$ test cases for step 1 of \cref{alg:blackCGP} and $\mathcal{M}_{\mathrm{T2}}$ consisting of $100$ test cases for step 2. 
Each test case contains the measurements resulting from $n_{\mathrm{s}}=10$ uncertainty trajectories with a length of $n_{\mathrm{k}}=n_{\mathrm{p}}+6$ time steps.
For the model order, we choose $n_{\mathrm{p}}=3$ for approximating the Lorenz model and $n_{\mathrm{p}}=2$ for the NARX1 model.
The accuracy of our black-box identification approach can be improved by using more complex models and more identification data. 
As such improvements always involve a trade-off in computation time, the decision will depend on the accuracy required for the application.

Next, we evolve a {population of model functions} with GP, using functionalities from the MATLAB toolbox GPTIPS2 \cite{searson2015gptips}.
The standard GP algorithm \emph{BlackGP} aims to minimize the least-squares errors for the training data over 100 generations for a population of 300 randomly initialized {model functions}. 
From the final population, we choose the {model function} with the highest fit to the data.
Additionally, we evaluate the CGP approach, denoted as \emph{BlackCGP}, where we minimize the least-squares cost function over $n_{\mathrm{g}}=95$ generations for $n_{\mathrm{f}}=300$ { model functions}, followed by $\tilde{n}_{\mathrm{g}}=5$ generations, where the cost $\ell_{\mathrm{C}}$ of \cref{eq:conf_opt} is minimized for $\tilde{n}_{\mathrm{f}}=100$ { model functions}. 

As the considered identification methods are of stochastic origin, the data generation and black-box identification process is repeated 30 times.
The normalized cost for all test suites is displayed in the boxplots in \cref{fig:nc} for the final models from all GP runs. 

While both approaches lead to similar normalized costs for the NARX1 model (see \cref{fig:nc_narx}), the costs for identifying the Lorenz system with \emph{BlackCGP} are much smaller (see \cref{fig:nc_nss}).
This is also shown in \cref{fig:narx,fig:nss} (we do not display the reachable sets $\mathcal{Y}_{\text{BlackCGP},k}$ and $\mathcal{Y}_{\text{BlackGP},k}$ at the time steps $k=n_{\mathrm{p}}+5$ and $k=n_{\mathrm{p}}+9$ in cases where they are too large): 
The models identified with \emph{BlackCGP} mimic the reachability of the true system well in an overapproximative way. 
In contrast, the results of \emph{BlackGP} are more volatile. 
The identified Lorenz model depicted in \cref{fig:nss} leads to more conservative reachable sets, while the reachable set of the identified NARX model depicted in \cref{fig:narx} is too small, as it does not contain all validation measurements at time step $k=n_{\mathrm{p}}+1$.
Furthermore, as can be seen from \cref{tab:comp}, the models identified with \emph{BlackGP} lead more often to failures in the conformance identification step. 
These failures result partially from exploding reachable sets when considering set-based inputs and partially from an insufficient influence of the input on the model dynamics, i.e., by increasing the scaling factors of the input set, we are not able to enlarge the reachable set in all dimensions, especially at early time steps, such that not all measurements can be enclosed and the model cannot be made reachset-conformant.

\subsection{Identification of a Vehicle Model}
Finally, we test our identification method with measurements of the automated vehicle EDGAR~\cite{karle2023edgar}, which is depicted in~\cref{fig:id}.
Given the limited number of available test cases, we only validate our white-box and gray-box identification methods.
We use the following kinematic model~\cite{rajamani2012vehicle}: 
\begin{align*}
    \dot{x}_t &= \begin{bmatrix}
        x_{t,4} \cos\bigl(\beta_t + x_{t,5}\bigr)\\
        x_{t,4} \sin\bigl(\beta_t + x_{t,5}\bigr)\\
        r_{\delta} u_{t,1} \\
        u_{t,2} \\
        x_{t,4} \cos\bigl(\beta_t\bigr) \frac{\tan(x_{t,3})}{l}
    \end{bmatrix} + u_{t,3:7}\\
    y_t &= \begin{bmatrix}x_{t,1}\\
    x_{t,2}\\
    \frac{1}{r_{\delta}}x_{t,3}\\
    x_{t,4}\\
    \end{bmatrix}  + u_{t,8:11},
\end{align*}   
where the slip angle $\beta_t$ can be computed as 
\begin{align*}
    \beta_t = \arctan\left (\tan\bigl(x_{t,3}\bigr) \frac{l_{\mathrm{r}}}{l}\right).
\end{align*}
The state vector $x_t=[p_{\mathrm{x},t}~p_{\mathrm{y},t}~\delta_{\mathrm{f},t}~v_{t}~\psi_{t}]^\top$ consists of the position in $x$ and $y$ direction, the steering angle at the front axle, the vehicle velocity, and the yaw angle, respectively. 
We obtain noisy measurements of the position, $y_{t,1}=\hat{p}_{\mathrm{x},t}$ and $y_{t,2}=\hat{p}_{\mathrm{y},t}$, the steering wheel angle $y_{t,3}=\hat{\delta}_{\mathrm{s},t}$, and the velocity $y_{t,4}=\hat{v}_t$.
The input and disturbance vector $u_t=[\dot{\delta}_{\mathrm{s},t}~a_{\mathrm{x},t}~w_t^\top~v_t^\top]^\top$ is composed of the steering wheel speed $\dot{\delta}_{\mathrm{s},t}$, the longitudinal acceleration $a_{\mathrm{x},t}$, the process noise vector $w_t$, and the measurement noise vector $v_t$.
The parameters $r_{\mathrm{\delta}}$, $l$, and $l_{\mathrm{r}}$ denote the transmission ratio from ${\delta}_{\mathrm{s}}$ to ${\delta}_{\mathrm{f}}$, the wheelbase, and the distance from the front axle to the center of gravity, respectively, and are given for the white-box identification and unknown for the gray-box identification.

The vehicle data is randomly split into an identification and a validation data set. 
The initial state estimate is computed from the measurements with $x_{*,0}=[\hat{p}_{\mathrm{x},0}~\hat{p}_{\mathrm{y},0}~r_{\delta}\hat{\delta}_{\mathrm{s},0}~\hat{v}_{0}~\hat{\psi}_{0}]^\top$
where the initial yaw angle is calculated as 
\begin{align*}
    \hat{\psi}_{0}=\arctan\left(\frac{\hat{p}_{\mathrm{y},1}-\hat{p}_{\mathrm{y},0}}{\hat{p}_{\mathrm{x},1}-\hat{p}_{\mathrm{x},0}}\right).
\end{align*}
From the $n_{\mathrm{m}}=84$ test cases of the identification data set with a length of $n_{\mathrm{k}}=6$ time steps and $n_{\mathrm{s}}=1$, we identify the following uncertainty sets with the white-box identification approach from~\cref{alg:white}:
\begin{align*}
\uncX^{(m)}&= \left\langle \mathbf{0}, \mathrm{diag}\left(\begin{bmatrix}
    0.00 & 0.55 & 0.00 & 0.66 & 0.00
\end{bmatrix}\right)\right\rangle, \\
{\mathcal{U}}_{\mathrm{in}} &= \left\langle  \mathbf{0}, \mathrm{diag}\left(\begin{bmatrix}
    0.00 & 0.00
\end{bmatrix}\right)\right\rangle, \\
{\mathcal{U}}_{\mathrm{w}} &= \left\langle  \mathbf{0}, \mathrm{diag}\left(\begin{bmatrix}
    0.03 & 0.02 & 0.00 & 0.06 & 0.00
\end{bmatrix}\right)\right\rangle, \\
{\mathcal{U}}_{\mathrm{v}} &= \left\langle  \mathbf{0}, \mathrm{diag}\left(\begin{bmatrix}
    2.82 & 1.84 & 0.05 & 0.79
\end{bmatrix}\right)\right\rangle,
\end{align*}
with $\uncU{k}^{(m)}={\mathcal{U}}_{\mathrm{in}} \times {\mathcal{U}}_{\mathrm{w}} \times {\mathcal{U}}_{\mathrm{v}}$.
The large values in ${\mathcal{U}}_{\mathrm{v}}$ indicate a large measurement uncertainty.
The sequential gray-box identification approach \emph{GraySeq} leads to larger initial state and process noise uncertainties. 
\begin{figure}[t]
        \centering        
        \includegraphics[width=0.6\linewidth, keepaspectratio, page=8]{figures/2024-02-14_NLIdentification_figure.pdf}
        
\subfloat[Position in $x$-direction.]{
        \centering        
        \begin{minipage}{.49\linewidth}
        \includegraphics[width=\linewidth, keepaspectratio]{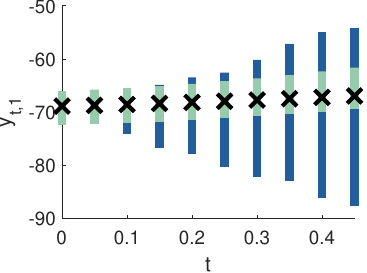}
        \end{minipage}
        \label{fig:edgar1}}
\subfloat[Position in $y$-direction.]{
        \centering        
        \begin{minipage}{.49\linewidth}
        \includegraphics[width=\linewidth, keepaspectratio]{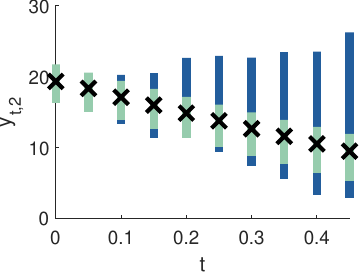}
        \end{minipage}
        \label{fig:edgar2}}

\subfloat[Steering wheel angle.]{
        \centering        
        \begin{minipage}{.49\linewidth}
        \includegraphics[width=\linewidth, keepaspectratio]{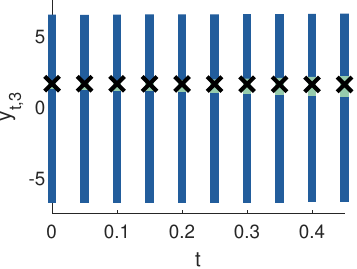}
        \end{minipage}
        \label{fig:edgar3}}
\subfloat[Velocity.]{
        \centering        
        \begin{minipage}{.49\linewidth}
        \includegraphics[width=\linewidth, keepaspectratio]{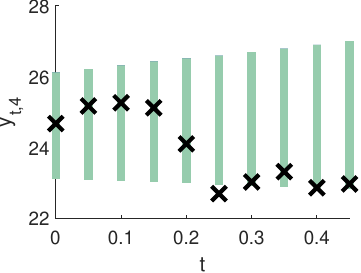}
        \end{minipage}
        \label{fig:edgar4}}
\caption{Reachable sets and measurement trajectories for a failure test case of the automated vehicle EDGAR.}
\label{fig:edgar}
\end{figure}

Next, we validate these results by computing the reachable sets for the $n_{\mathrm{m}}=304$ test cases of the validation data set with a length of $n_{\mathrm{k}}=10$ time steps. 
98.9\% of the measurements are contained in their corresponding reachable set for the white-box model and 93.6\% for the gray-box model, which shows that the proposed identification framework works well for real-world data.
One of the failure test cases is visualized in~\cref{fig:edgar}. We can see in~\cref{fig:edgar4} that the measurement at time $0.25s$ is marginally outside the reachable set (the reachable velocities are identical for both models for this test case).
Introducing safety factors for scaling the identified uncertainty sets decreases the failure rate, i.e., a safety factor of $\epsilon=1.2$ leads to 99.2\% contained measurements for the white-box model and 97.6\% for the gray-box model, while all measurements are contained in the respective reachable sets for both models when using a safety factor of $\epsilon=3$. 
The magnitude of the necessary safety factor shows that the data contains some outliers (measurements that deviate substantially from the expected outputs).
Thus, a lower failure rate could also be achieved by detecting and removing outliers from the measurement data.


\section{Conclusions}
\label{sec:conclusions}

We present a novel framework for identifying reachset-conformant models for linear and nonlinear system dynamics, adaptable to various levels of prior knowledge.
A pivotal contribution is the introduction of the {GLO approximation}, which unifies reachability analysis across different model types.
Leveraging this {approximation}, we introduce the first reachset-conformant identification method capable of handling potentially non-additive uncertainty sets and input-output models. 
Our framework also extends to the estimation of unknown model parameters and the identification of reachset-conformant black-box models using genetic programming, allowing for modeling complex systems with minimal prior information.
Through extensive numerical experiments involving the identification of linear state-space, ARX, nonlinear state-space, and NARX models, as well as the modeling of an automated vehicle using real-world data, we demonstrate the robustness and practical applicability of our proposed approaches. 
Across all experiments, we successfully identify reachset-conformant models, whose reachable sets tightly enclose all measurements, within competitive computation times.

While our work guarantees reachset conformance for all trajectories considered in the identification process, we did not make any mathematical statements about the generalizability of the results to unseen data.
Although our experiments demonstrate promising performance of the identified models on unseen data, future research will focus on deriving probabilistic guarantees, potentially utilizing the scenario approach \cite{calafiore2006scenario}. 
Despite these limitations, our approach offers a valuable tool for enhancing reliability and mitigating conservatism in the formal verification and control of real-world systems. 
It represents a significant step towards transferring model-based verification results to practical applications, thereby addressing a critical gap in current methodologies.


\bibliographystyle{ieeetr}
\vspace{0.1cm}
\section*{References}
\vspace{-0.6cm}
\bibliography{literature}

\begin{IEEEbiography}[{\includegraphics[width=1in,height=1.25in,clip,keepaspectratio]{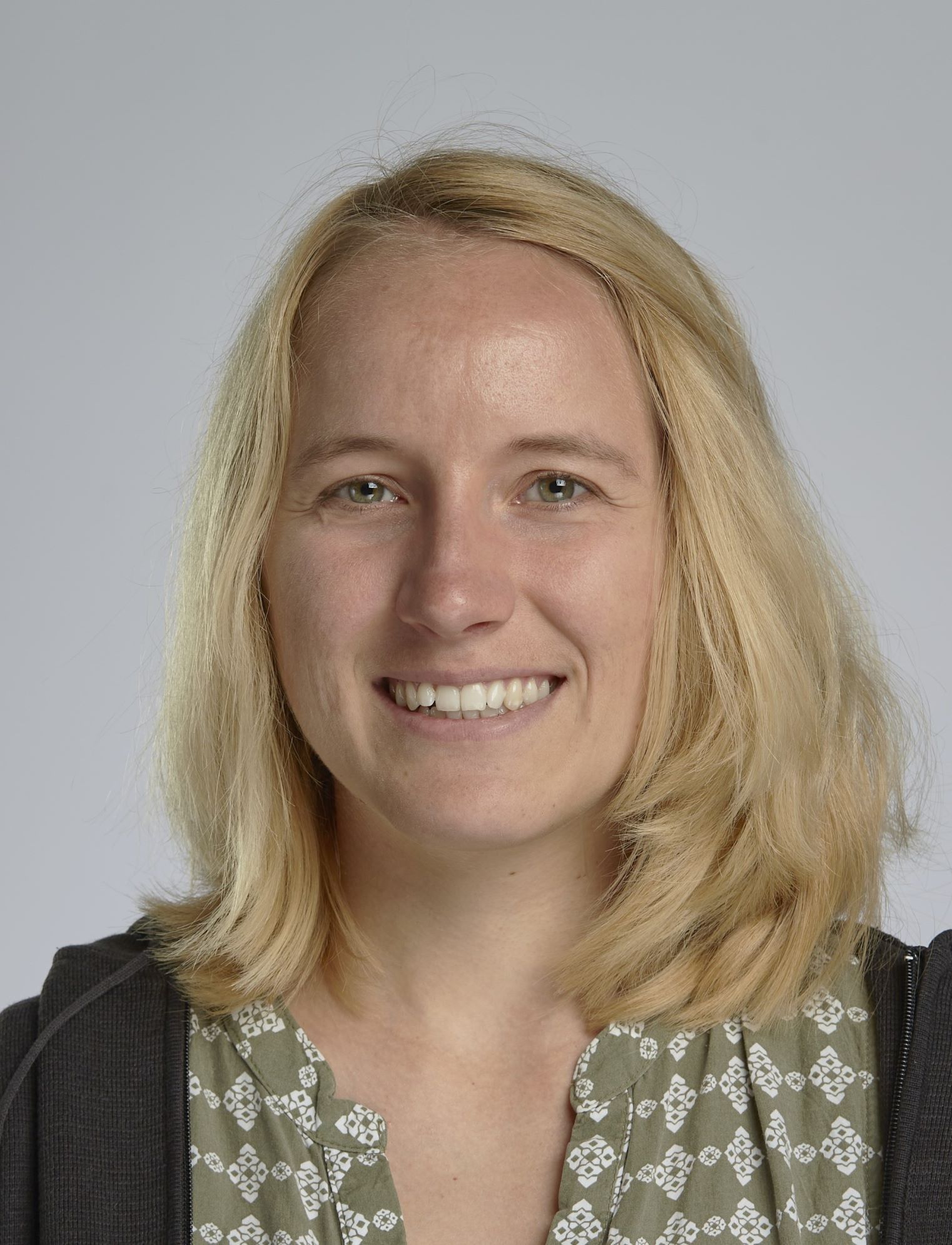}}]{Laura Lützow} received the Bachelor of Science degree in mechatronics from the Technical University of Ilmenau, Germany, in 2020, and the Master of Science degree in robotics, cognition and intelligence from the Technical University of Munich, Germany, in 2022.
In 2022, she was a visiting student researcher at
the Massachusetts Institute of Technology, USA.
Currently, she is working toward a Ph.D. degree in computer science with the Cyber-Physical Systems Group, School of Computation, Information and Technology, at the Technical University of Munich.
Her research interests include system identification, control theory, and reachability analysis, with applications to safety-critical systems.
\end{IEEEbiography}

\begin{IEEEbiography}
[{\includegraphics[width=1in,height=1.25in,clip,keepaspectratio]{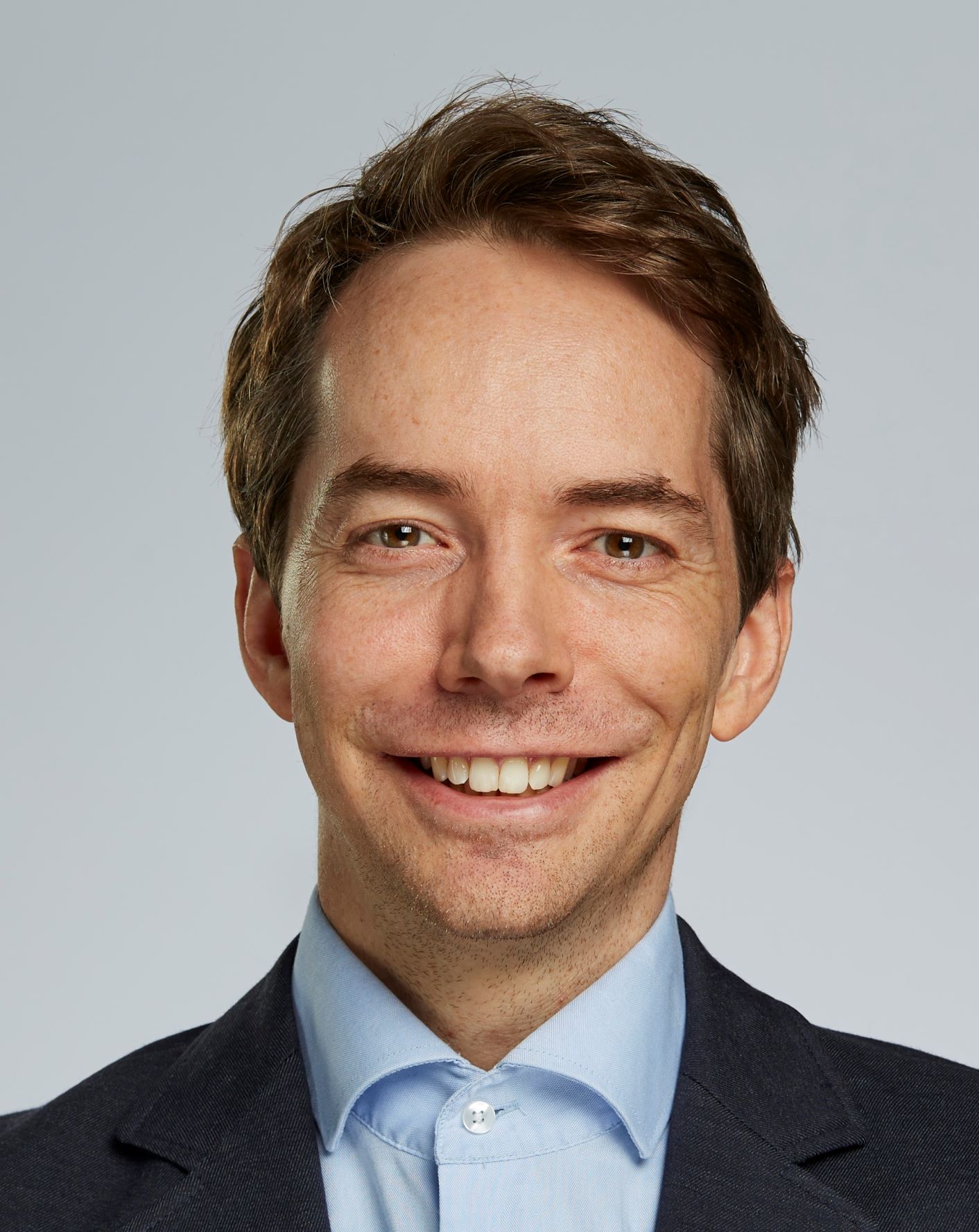}}]{Matthias Althoff} (Member, IEEE)
received the Diploma Engineering degree in mechanical engineering and the Ph.D. degree in electrical engineering, both from the Technical University of Munich, Germany, in 2005 and 2010, respectively.
From 2010 to 2012, he was a Postdoctoral Researcher at Carnegie Mellon University, Pittsburgh, PA, USA, and from 2012 to 2013, he was an Assistant Professor at the Technical University of Ilmenau, Germany.
He is currently an Associate Professor in computer science with the Technical University of Munich. His research interests include formal verification of continuous and hybrid systems, reachability analysis, planning algorithms, nonlinear control, robotics, automated vehicles, and power systems.
\end{IEEEbiography}

\end{document}